%% file: main.tex
\documentclass{article}
\usepackage{a4wide}
\usepackage{dcolumn}
\newcolumntype{d}{D{.}{.}{-1}}
\usepackage{subfigure}
\usepackage{fancyvrb}
\usepackage{caption}
\usepackage{graphicx}
\fvset{fontsize=\footnotesize,xleftmargin=2em}
\input abbrevs

\title{ON TECHNIQUES FOR BARELY COUPLED MULTIPHYSICS}

\author{Rainald L\"ohner$^1$, Harbir Antil$^2$ and Sebastian Sch\"ops$^3$ \\[5pt]
  {\itshape $^1$Center for Computational Fluid Dynamics} \\
  {\itshape ~George Mason University, Fairfax, VA 22030-4444, USA} \\
  {\itshape $^2$Center for Mathematics and Artificial Intelligence} \\
  {\itshape ~George Mason University, Fairfax, VA 22030-4444, USA} \\
  {\itshape $^3$Computational Electromagnetics Group} \\
  {\itshape ~TU Darmstadt, Darmstadt, Germany } \\
}

\begin{document}

\maketitle

\begin{abstract}
A technique to combine codes to solve barely coupled multiphysics problems
has been developed. Each field is advanced separately until a stop is
triggered. This could be due to a preset time increment, a preset number
of timesteps, a preset decrease of residuals, a preset change in unknowns,
a preset change in geometry, or any other physically meaningful quantity.
The technique allows for a simple implementation in coupled codes using
the loose coupling approach. 
Examples from evaporative cooling of electric motors, a problem that
has come to the forefront with the rise of electric propulsion in the 
aerospace sector (drones and air taxis in particular) shows the
viability and accuracy of the proposed procedure. 
\end{abstract}

\section{INTRODUCTION}
Over the last five decades computational mechanics has matured rapidly.
In each of the core disciplines -~fluid dynamics, structural dynamics, 
combustion, heat transfer, acoustics, electromagnetics, mass transfer,
control, etc.~- robust and efficient numerical techniques have been
developed, and a large code base of academic, open source and commercial
codes is available. The acquisition of many of these commercial codes by
the leading CAD-vendors attests to the desire to streamline the typical
computational mechanics workflow (CAD, definition of boundary conditions, 
loads and physical parameters, solution, mesh adaptation, post-processing)
by integrating all parts into a single application. \\
The ability to obtain accurate and timely results in each of the core
disciplines or metiers has 
prompted the desire to reach the same degree of simplicity in computing
multi-physics problems. Considering the different approximation levels
possible in each metier, any multidisciplinary capability must have 
the ability to quickly switch between approximation levels, abstractions,
models, grids, and, ultimately, codes. It is clear that only those 
approaches that allow a maximum of flexibility, i.e.:
\begin{itemize}
\item[-] Linear and nonlinear metier-specific models adapted to the
problem at hand;
\item[-] Different, optimally suited discretizations for each discipline;
\item[-] Modularity and extendibility in models and codes;
\item[-] Fast multidisciplinary problem definition; and
\item[-] Fully automatic grid generation for arbitrary geometrical complexity
\end{itemize}
will survive in the long term. 

A large class of coupled problems exhibits large disparity of timescales.
Examples include evaporative cooling (where the flowfield may be established
in seconds while the temperature field requires minutes), sedimentation of
rivers and estuaries (where the flowfield is established in seconds while
the filling up of a channel takes weeks), deposition of cholesterol in
arteries (where the flowfield is established after two heartbeats while
the deposition can take years), the wear of semi-autogenous grinding (SAG)
mills (where the movement of steel balls and mineral-rich rocks and mud
is established in minutes while the wear of the liners can take hours), and
many others. We denote this class of problems as `barely coupled'.
In each of these cases a coupling is clearly present.
However, due to physics and nonlinear effects one can not simply run with
a fully coupled time discretization using very large timesteps. This would
lead to incorrect results. On the other hand, it becomes very costly 
(and in many cases impossible due to constraints in computing time) 
to run in a strictly time-accurate manner. The recourse advocated 
here is to run each problem in a controlled way to a quasi steady-state, 
and to couple the different disciplines in a loose manner. 
This approach also offers the possibility of easily coupling 
different software packages (in-house, open source or commercial).

\section{BARELY COUPLED TIMESTEPPING}
The two most popular ways of
marching in time when coupling different codes are the so-called
loose \cite{Loh95,Ceb97,Loh98,Ceb05,Bun06,Yan06,Sot10,Mic11,Per19} and 
tight \cite{Bun06,Kue08,Sch08,Cle12}
coupling. Both are shown diagrammatically in Figure~2.1, where the
term `app' stands for application or physics/code.

\vbox{
\vskip 18pt
\centerline{
\includegraphics[width=14.0cm]{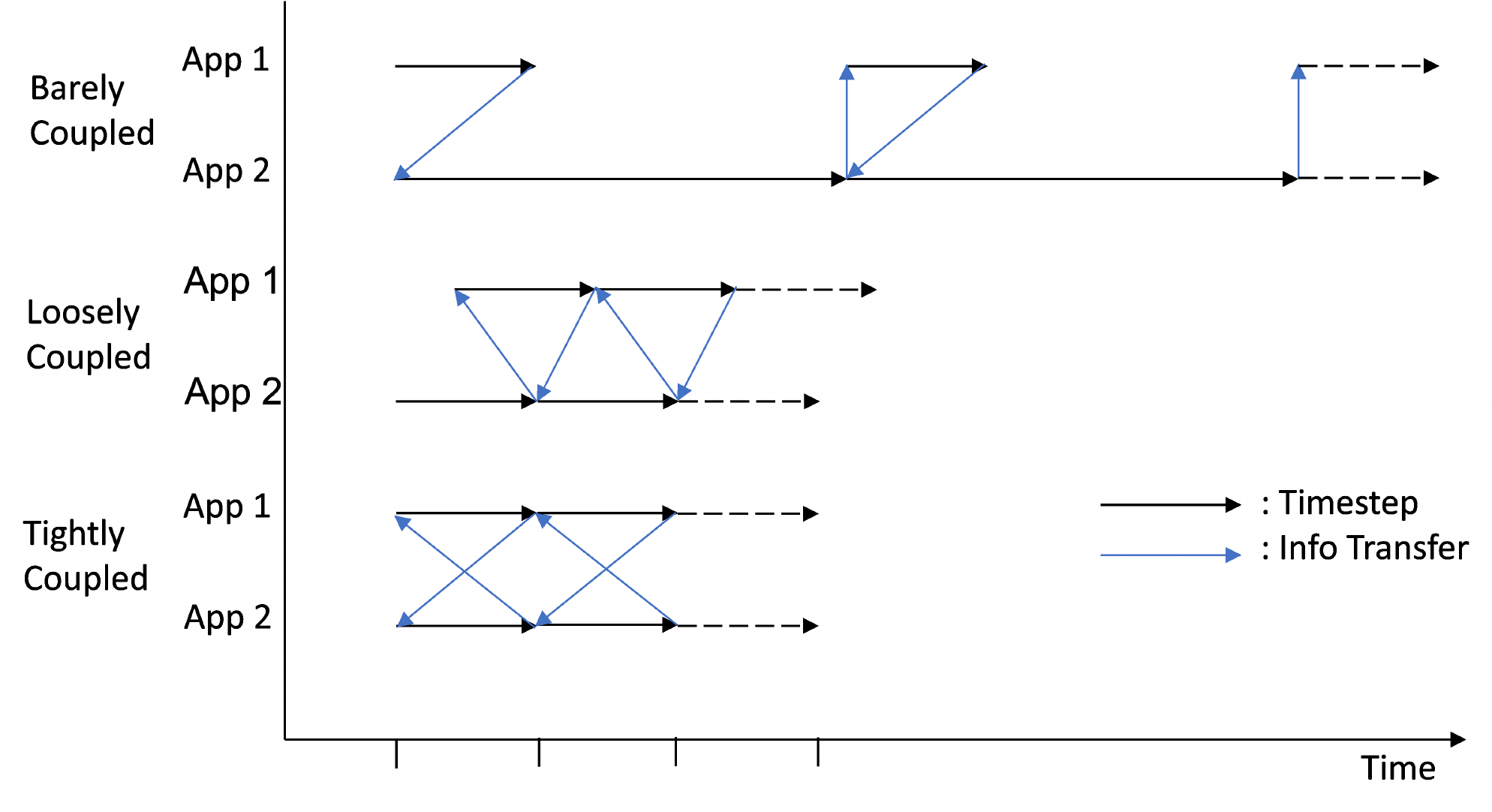}
}
\vskip 5pt
\centerline{Figure 2.1: Advancement of Two Coupled Codes in Time}
\vskip 10pt
}

\par \noi
Loose coupling advances each code independently until the time of
each code exceeds the time of the other code(s). Once an individual
code reaches this time, the information is passed to the other code(s)
and they advance in turn. Loose coupling has been used extensively for
fluid-structure and fluid-structure-thermal interaction, particularly
when explicit timestepping is required for one of the applications due
to the physics \cite{Loh95,Sot10,Mic11,Per19}.
\par \noi
Tight coupling, on the other hand, seeks to synchronize in a precise way
the timesteps and the associated information transfer between the codes.
Within each timestep, the codes are advanced several times until the
iteration converges to an equilibrium state at the end of the timestep
\cite{Bun06,Kue08,Sch08,Cle12}. 
\par \noi
Both of these procedures are suboptimal for applications where the timescales
at which the physically relevant processes occur are vastly different, yet
a detailed, time-accurate simulation is required for any of the codes.
There are numerous applications where this is the case: in fact, one could
surmise that there are many more applications that fall under this category
than those amenable to the previously mentioned forms of coupling.
What all of these applications have in common
is the establishment of a quasi-steady/quasi-periodic pattern in one of
the codes that affects the much slower evolution of the other code(s).
Running all codes throughout the complete time required is neither
efficient, nor (and more importantly): necessary~(!). Which results in
a new form of coupling: {\bf barely coupled} multiphysics.
The code with the `fast transient' is run until a quasi-steady/quasi-periodic
pattern is observed. This information (which may be transient data, i.e.
not just a simple time-averaged field) is then passed to the `slow transient'
code(s). This code then advances in time until the geometry, the physics, or
any other measure that would exceed an error tolerance is reached. The
information at this time is then passed to the `fast transient' code.
The `fast transient' code then continues from this point in time, i.e.
this code {\bf ignores all the intermediate time that has passed since the
last update}. The sequence is shown at the top of Figure~2.1. \\
The implementation of barely coupled timestepping may be summarized as 
follows:
\par \noi
For Each Iteration/Timestep:
\begin{itemize}
\item[-] Do: Loop Over the Codes Used
\item[-] Import All Relevant Information:
\item[ ] Updated Geometry, Velocities, Temperatures, 
\item[ ] Forces, Heat Fluxes, ...
\item[-] Run Code to Completion/Quasi Steady-State
\item[-] Export All Relevant Information
\item[ ] Forces, Heat Fluxes, 
\item[ ] Updated Geometry, Velocities, Temperatures, ...
\item[-] EndDo: Loop Over the Codes Used
\end{itemize}
\par \noi
The implicit assumption made here is that each individual code only
imports and exports information from a coupling code or library that
does all the required interpolation/projection operations. \\
The most important research question is how to detect the completion
of each physics portion/code. Options that immediately come to mind
include:
\begin{itemize}
\item[-] Preset time increments (this could still imply many timesteps);
\item[-] Preset decrease of residuals (useful for steady state cases);
\item[-] Preset quasi-periodic behaviour (useful for cyclic loads);
\item[-] Preset energy balance / tolerance;
\item[-] Preset change in unknowns; and
\item[-] Preset change in geometry.
\end{itemize}

\section{EVAPORATIVE COOLING OF ELECTRICAL MOTORS}
The modeling of evaporative cooling is a good example for barely coupled
multiphysics: the high-fidelity modeling requires the solution of the
Maxwell equations to obtain the heat sources, a mix of continuum and 
particle methods for the fluids (which reaches a quasi steady-state in less
than a second), as well as a heat conduction solver for the rotor and 
stator (where the quasi-steady temperature field establishes itself in
minutes). The problem has come to the forefront with the rise of
electric propulsion in the aerospace sector (drones and air taxis in 
particular). Optimization of electrical motors (miniaturization, weight
reduction, increase efficiencies) in some cases has led to overheating.
Evaporative cooling is a possibility that has been actively 
pursued \cite{Gro21}. The PDEs and ODEs used to describe and compute the 
multiphase flow- and heat-fields have been summarized in Appendices~1,2. \\


In the sequel we consider two scenarios for electrical motors that 
are cooled evaporatively via particles that are injected in the narrow 
gap between the rotor
and the stator. This is a typical barely coupled multiphysics case:
the flowfield in the gap is established in less than a second, while the
heating of the motor can take minutes. \\

\bs \noi
3.1 \ub{Simple Gap Region}

\ms \noi
The geometry chosen, as well as the general boundary conditions, are shown
in Figure~3.1a.
The physical parameters were set as follows (all units in cgs):

\par \noi
Geometry of the section computed:
\begin{itemize}
\item[-] Inner radius: 2.0
\item[-] Outer radius: 2.1
\item[-] Outermost radius: 4.0
\item[-] Depth(z): 2.0
\end{itemize}
\par \noi
Flow:
\begin{itemize}
\item[-] Inflow: air: 100cm/sec, te=300K, dens=0.00122, p=1.0e6
\item[-] Rotation of inner part: 600 rpm
\item[-] Viscosity: 0.1850E-03
\item[-] Conductivity: 0.2400E+04
\item[-] Specific heat coefficient: 0.1000E+08
\item[-] Particles: water, d=0.01
\end{itemize}
\par \noi
Solid/Heat:
\begin{itemize}
\item[-] Density: 7.85
\item[-] Specific heat coefficient: 420.0e+4
\item[-] Conductivity: k=50.0e+5
\item[-] Volumetric heat load: 2.0e+7
\end{itemize}

\par \noi
The discretizations used are shown in Figures~3.1b,c. Note that optimal
discretizations have been employed for each field, and the grids do
not match at the interface. Based on previous scoping runs, during each
barely coupled step the flowfield 
was advanced for 500~timesteps with a timestep of approximately
$\Delta t = 0.3 \cdot 10^{-4}$, which corresponds to an advective 
Courant-number of $C=0.8$. This number of timesteps is required 
for the flowfield to reach a quasi steady state.
The temperature field, on the other hand, was advanced for 10~timesteps
with a timestep of $\Delta t = 10$.
Figures~3.1d-f show the temperature field 
obtained, the velocity in the gap, the velocity of the flowfield and
the particles, as well as the temperature of the flow and the particles.
Note that evaporation of particles does occur.
Figures~3.1g-i show the sum of the differences in temperature at the 
interface
between the fluid and solid domains, the heat loads seen by each of the
domains (CFD: computational fluid dynamics, i.e. the flow domain,
CTD: computational thermo-dynamics, i.e. the solid region)
in the gap region, and the minimum and maximum temperature seen
on the gap surface by the two codes/domains. 
Note the `transient beginning', indicative of the
strong coupling of the problem at hand, and the subsequent `settling to
steady state'.

\vbox{
\vskip 18pt
\cl{
\includegraphics[width=12.0cm]{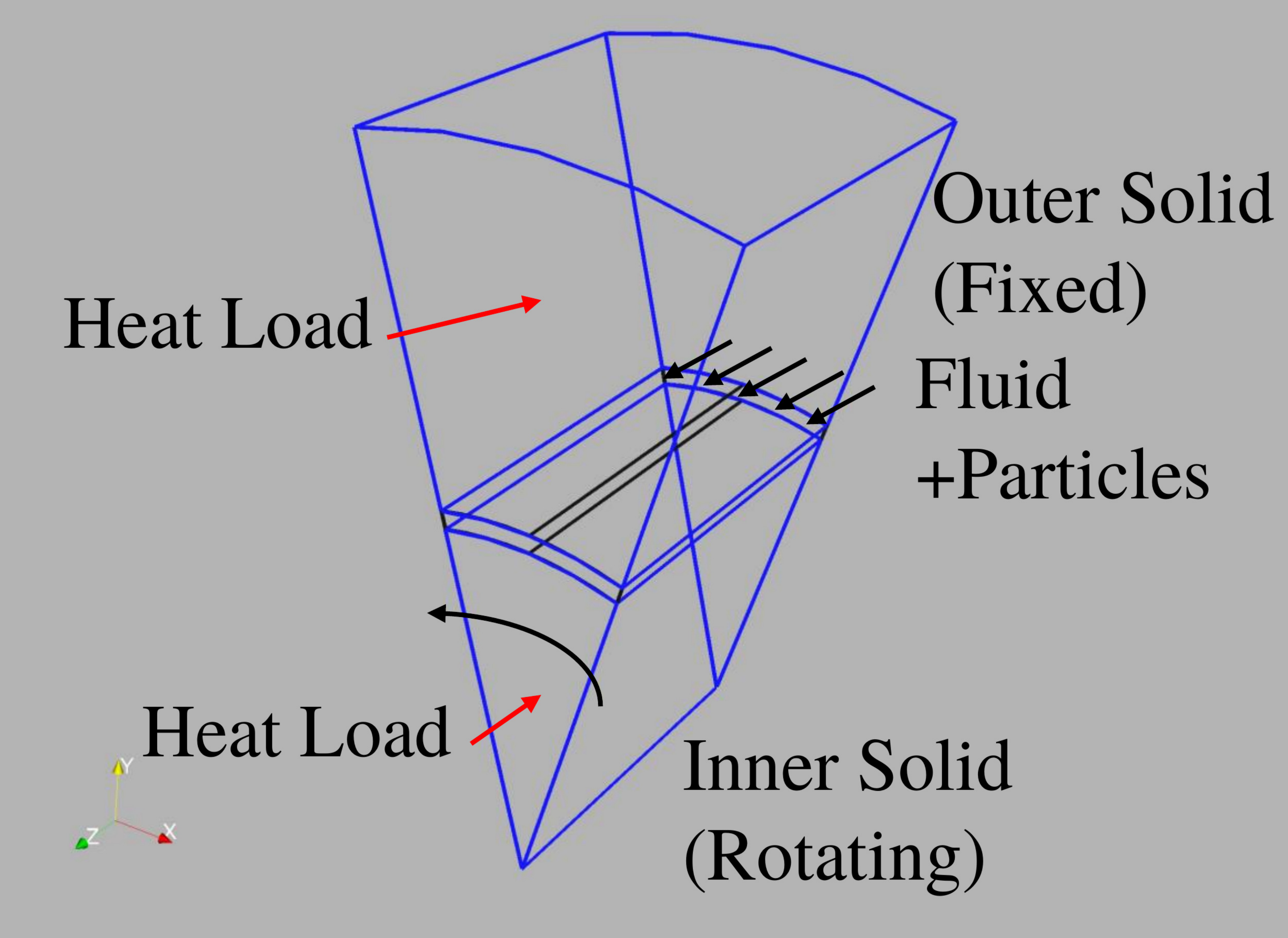}
}
\vskip 5pt
\cl{Figure 3.1a~~E-Motor: Problem Definition}
}

\vbox{ 
\vskip 18pt
\cl{
\includegraphics[width=12.0cm]{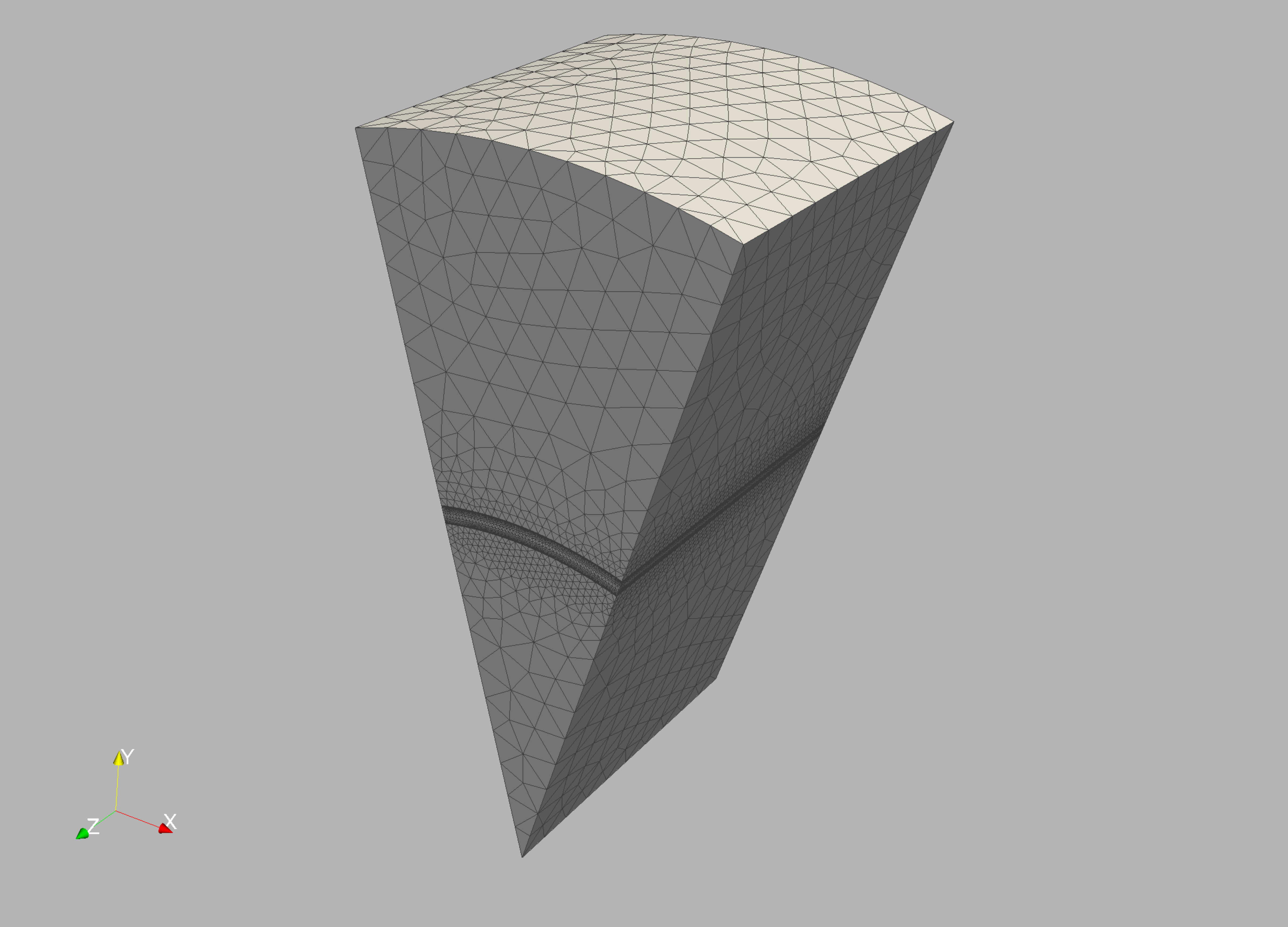}
}
\vskip 5pt
\cl{Figure 3.1b~~E-Motor: Grids Used}
}

\vbox{
\vskip 18pt
\cl{
\includegraphics[width=12.0cm]{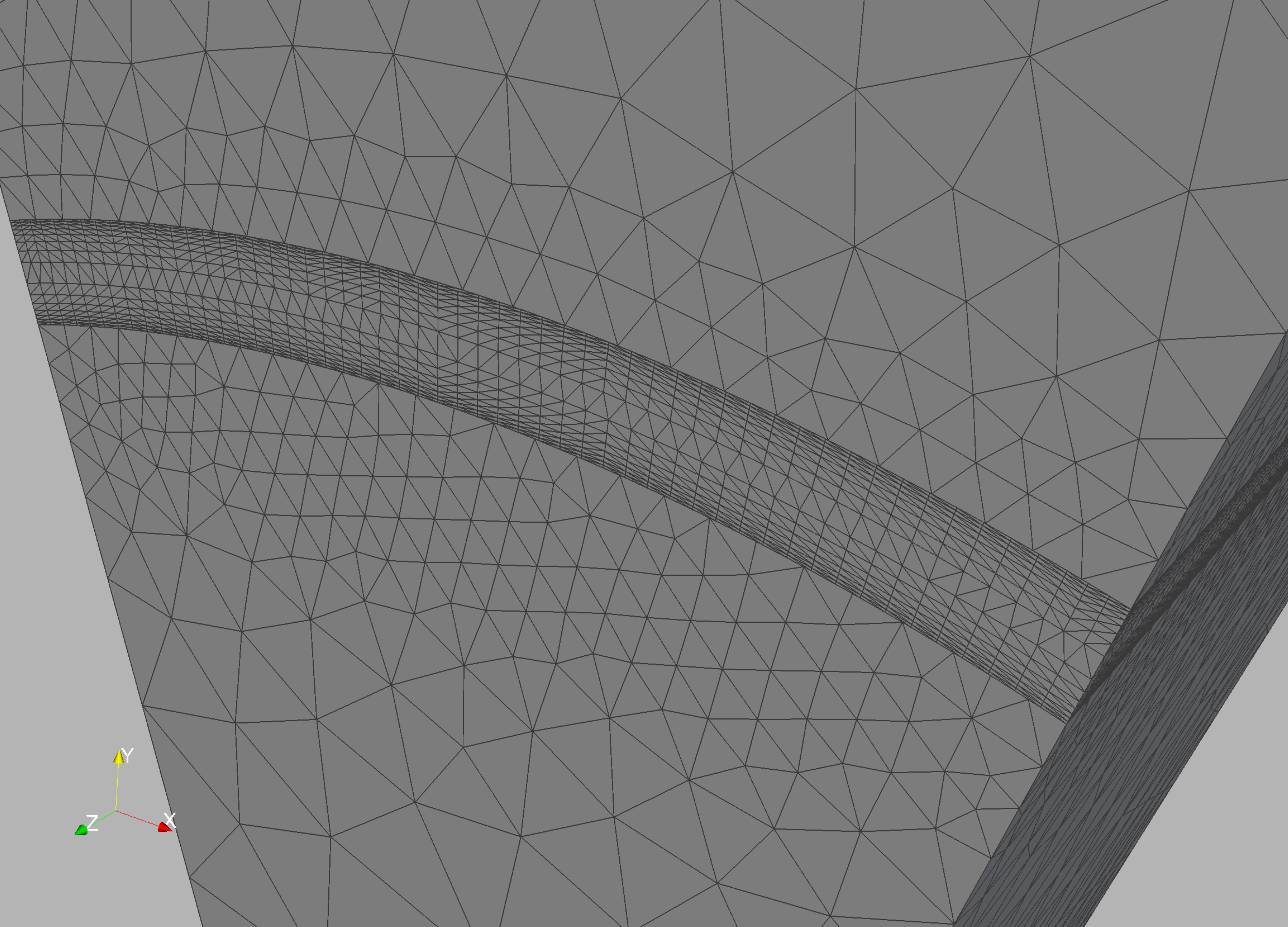}
}
\vskip 5pt
\cl{Figure 3.1c~~E-Motor: Grids Used (Detail)}
}

\vbox{
\vskip 18pt
\cl{
\includegraphics[width=12.0cm]{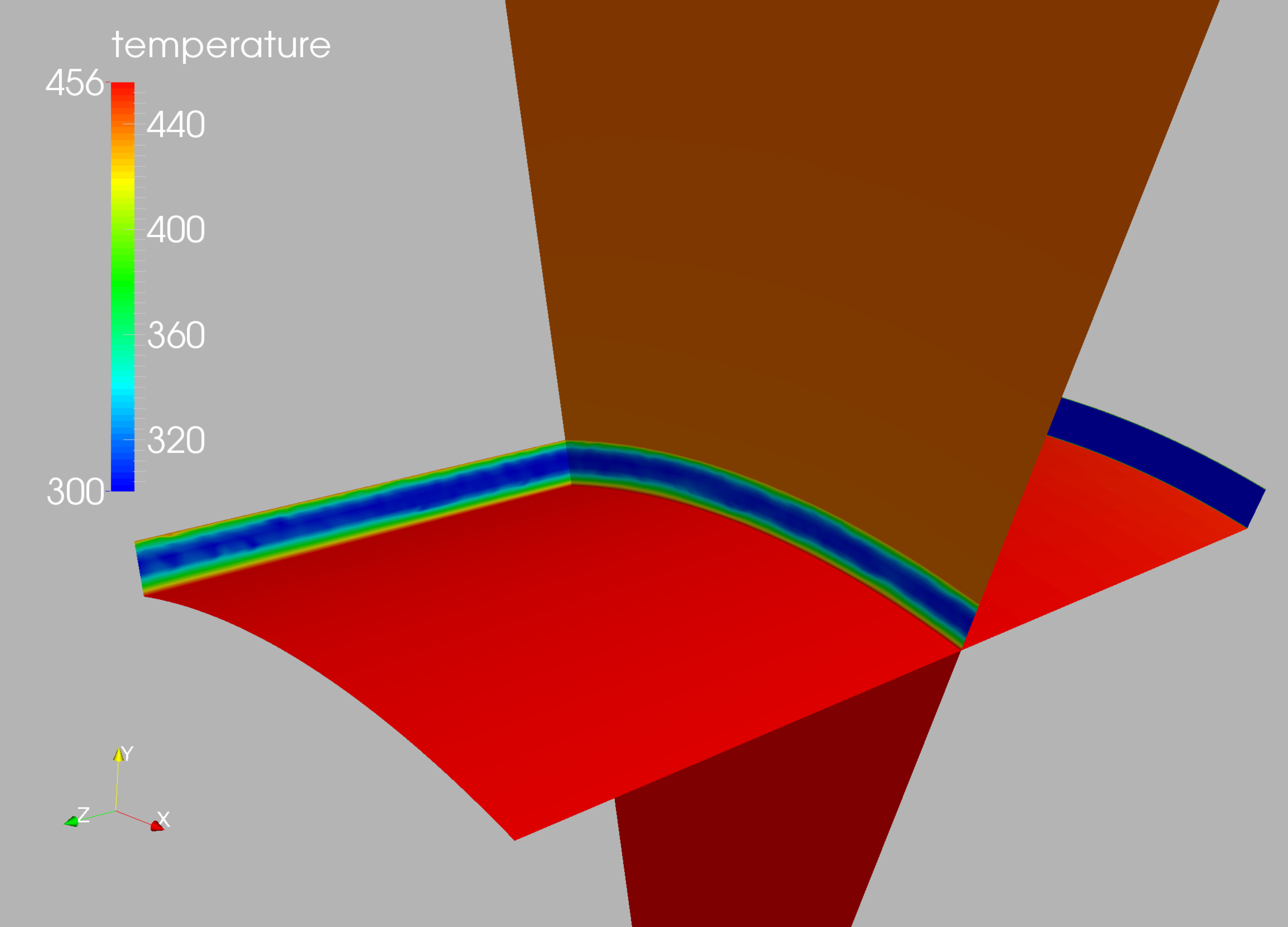}
}
\vskip 5pt
\cl{Figure 3.1d~~E-Motor: Temperature Field Obtained}
}

\vbox{
\vskip 18pt
\cl{
\includegraphics[width=12.0cm]{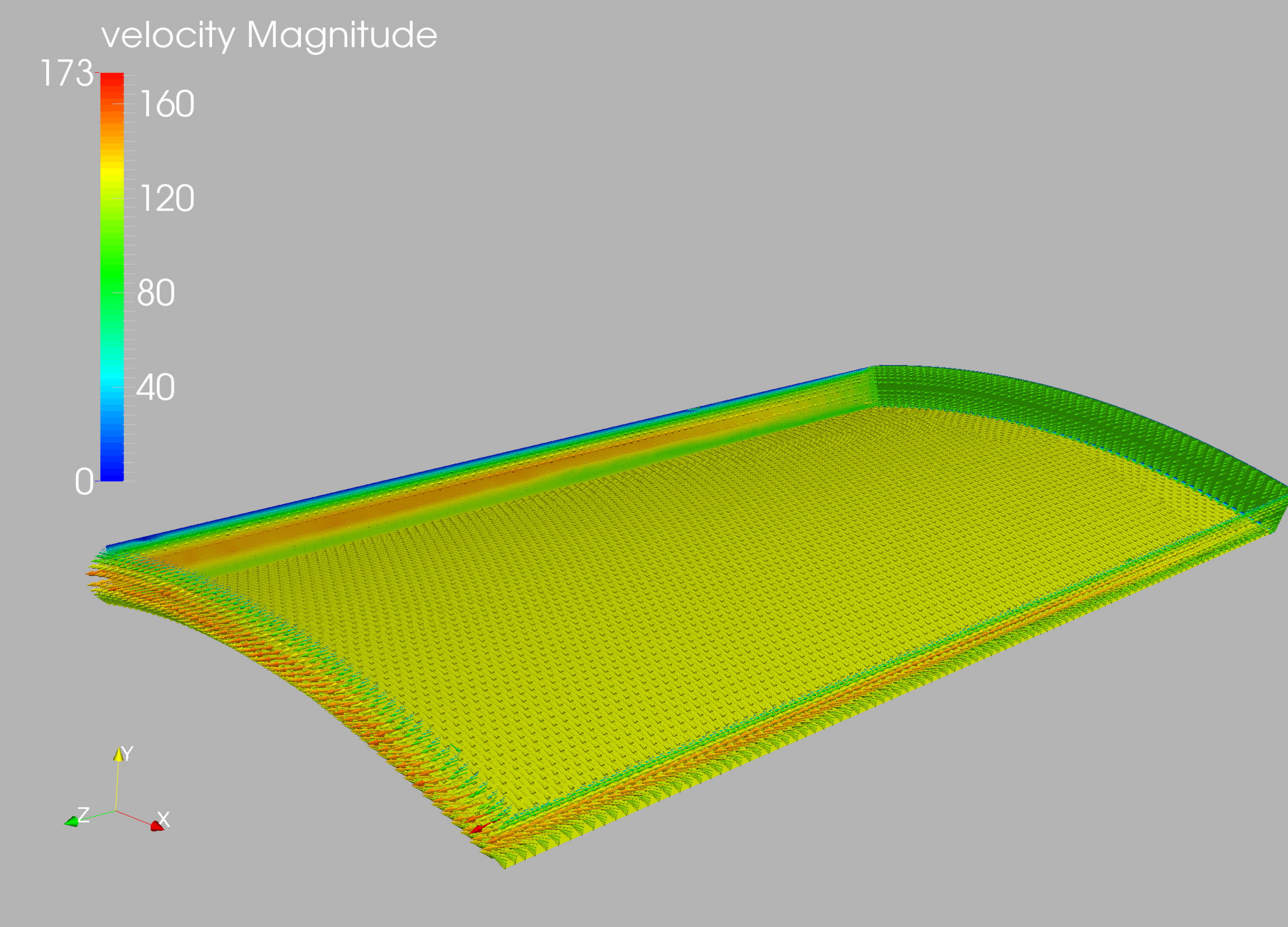}
}
\vskip 5pt
\cl{Figure 3.1e~~E-Motor: Velocity Field Obtained}
}

\vbox{
\vskip 18pt
\cl{
\includegraphics[width=12.0cm]{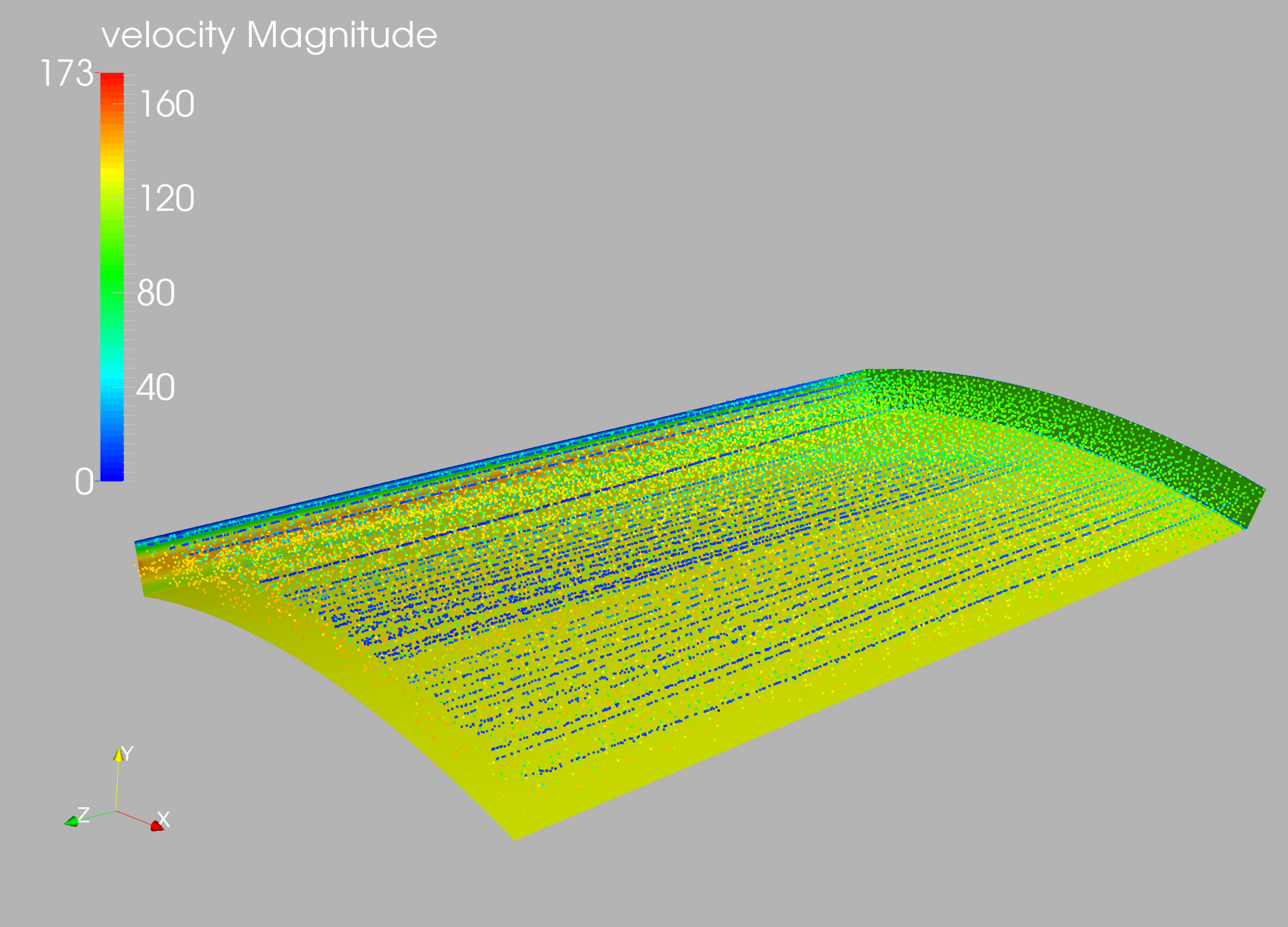}
}
\vskip 5pt
\cl{Figure 3.1f~~E-Motor: Velocity Field of Gas (Surface) and Particles (Volume)}
}

\vbox{     
\vskip 18pt
\cl{
\includegraphics[width=12.0cm]{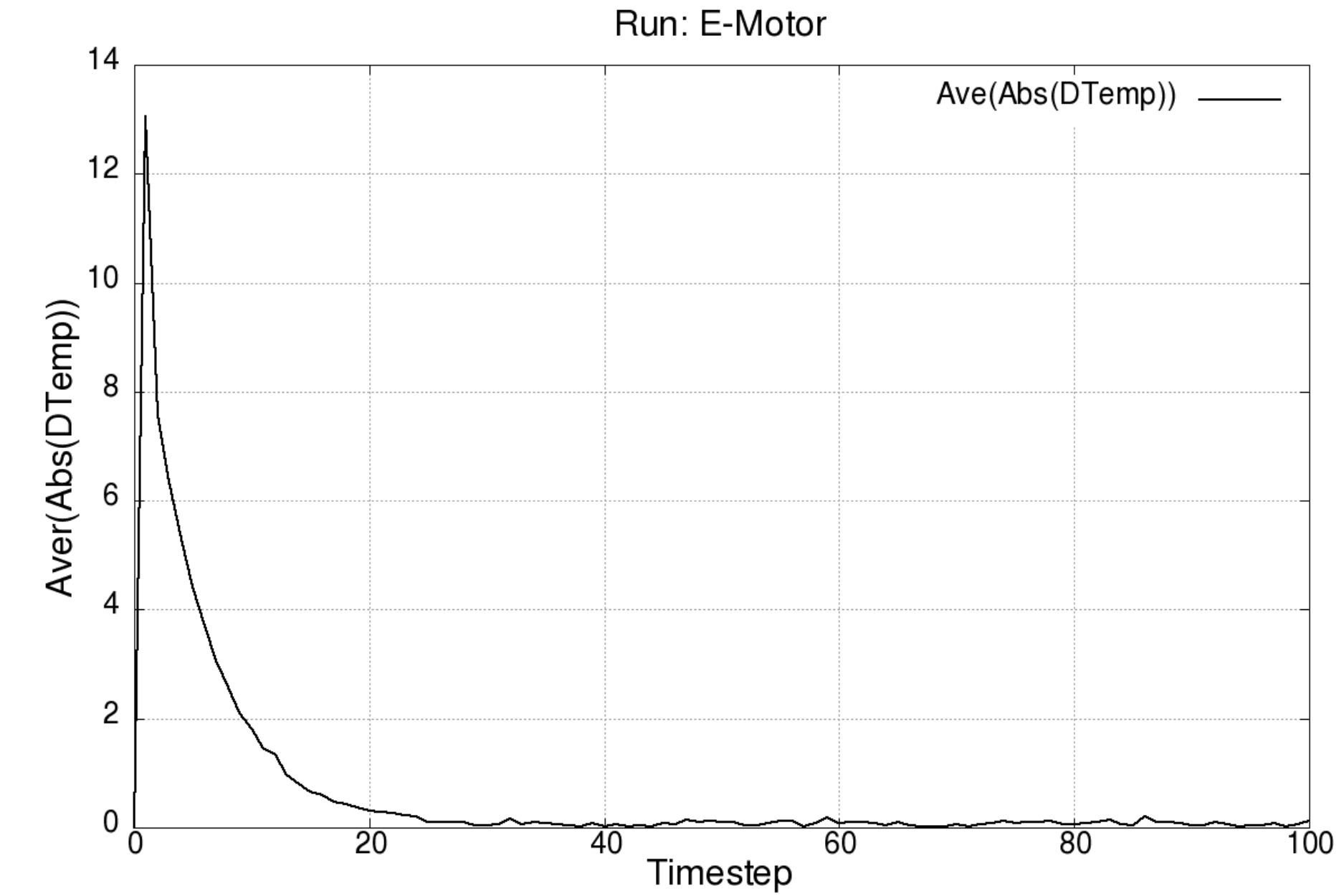}
}
\vskip 5pt
\cl{Figure 3.1g~~E-Motor: Difference in Surface Temperatures Between Flow
and Heat Solvers at the Surface of the Gap}
}

\vbox{
\vskip 18pt
\cl{
\includegraphics[width=12.0cm]{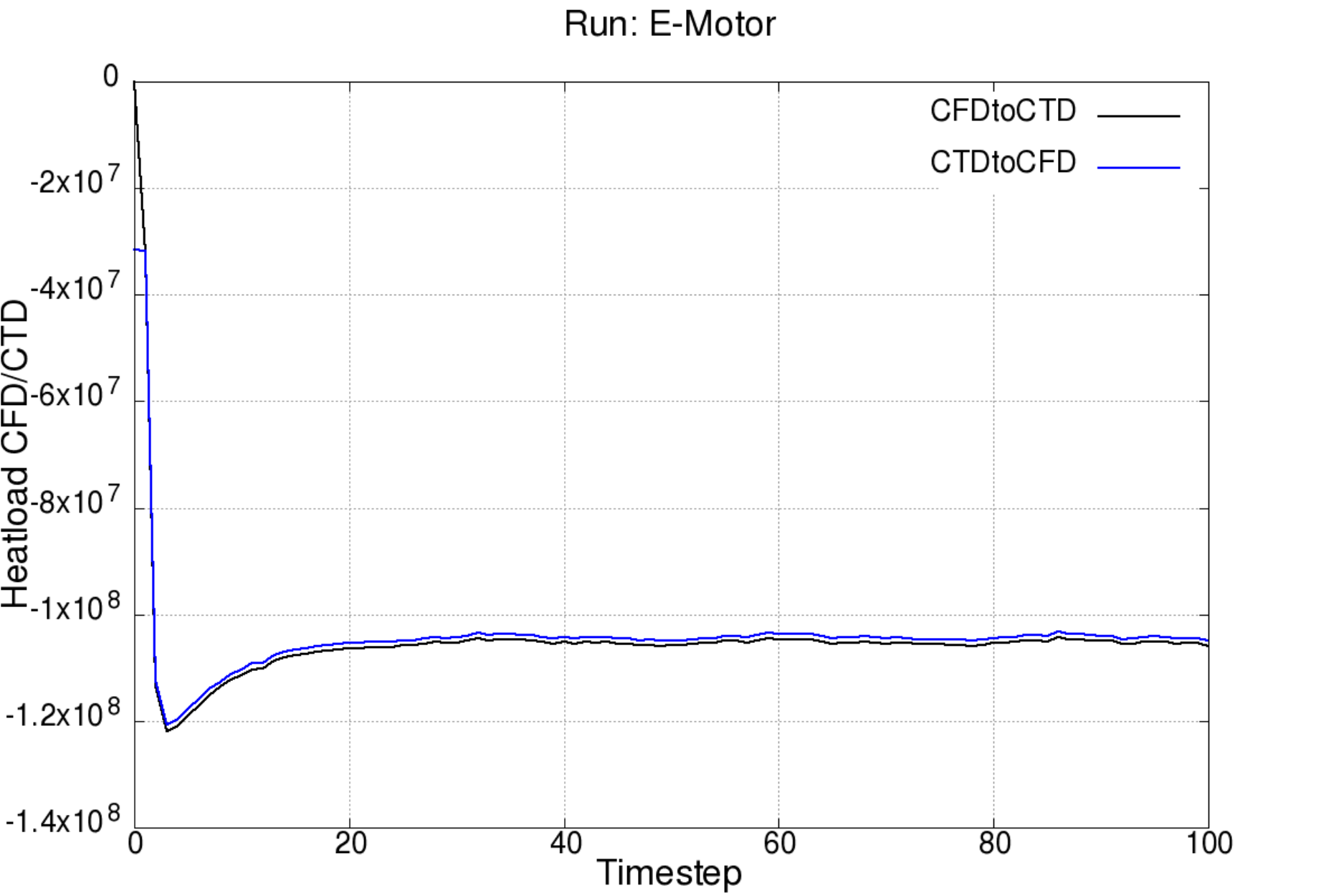}
}
\vskip 5pt
\cl{Figure 3.1h~~E-Motor: Heat Loads Seen by Flow 
and Heat Solvers at the Surface of the Gap}
}

\vbox{
\vskip 18pt
\cl{
\includegraphics[width=12.0cm]{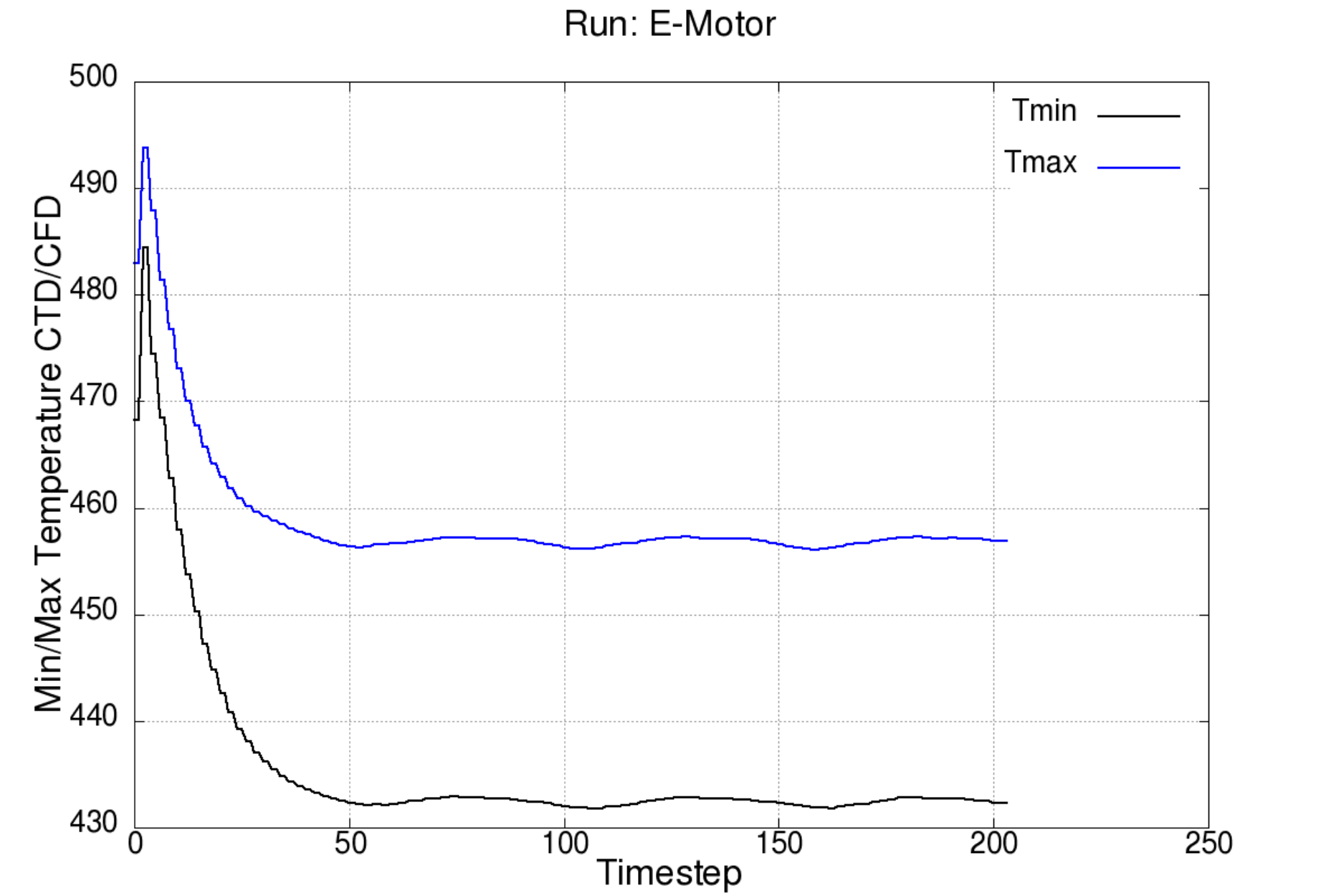}
}
\vskip 5pt
\cl{Figure 3.1g~~E-Motor: Minimum and Maximum Temperature 
at the Surface of the Gap}
}

\bs \noi
3.2 \ub{Complete Motor}

\ms \noi
The geometry chosen is shown in Figure~3.2a,b, and is taken from
the repository than be accessed under \cite{Hei24}. Notice the arrangement
of copper coils on both sides of the stator. In this case the particles
are injected in the (axial) middle of the gap, then travel in the gap
towards both ends, and are then injected with a centrifugal direction into
the flow. The aim is for the particles to remove heat both in the
gap and as they come into contact with the coils.
The physical parameters were set as follows (all units in cgs):

\par \noi
Geometry of the section computed:
\begin{itemize}
\item[-] Inner radius: 6.240
\item[-] Outer radius: 6.285
\item[-] Outermost radius: 10.0
\item[-] Length: 10.0
\item[-] Radius of Casing: 13.0
\end{itemize}
\par \noi
Flow:
\begin{itemize}
\item[-] Inflow: air: 100cm/sec, te=300K, dens=0.00122, p=1.0e6
\item[-] Rotation of inner part: 600 rpm
\item[-] Viscosity: 0.1850E-03
\item[-] Conductivity: 0.2400E+04
\item[-] Specific heat coefficient: 0.1000E+08
\item[-] Particles: water, d=0.01
\end{itemize}
\par \noi
Solid/Heat: Magnets
\begin{itemize}
\item[-] Density: 7.85
\item[-] Specific heat coefficient: 420.0e+4
\item[-] Conductivity: k=50.0e+5
\end{itemize}
Solid/Heat: Copper Coils in Stator and Outside Stator
\begin{itemize}
\item[-] Density: 8.94
\item[-] Specific heat coefficient: 385.0e+4
\item[-] Conductivity: k=385.0e+5
\end{itemize}
Solid/Heat: Steel for Rotor, Shaft and Stator
\begin{itemize}
\item[-] Density: 7.85
\item[-] Specific heat coefficient: 420.0e+4
\item[-] Conductivity: k=50.0e+5
\end{itemize}
Heat Loads:
\begin{itemize}
\item[-] Volumetric Heat for Steel in Rotor: 2.0e+7
\item[-] Volumetric Heat for Steel in Stator: 1.0e+6
\end{itemize}

\par \noi
The discretizations used are shown in Figures~3.2c-e. The solid was
discretized with about 4.3~Mtets, while the fluid field exceeded
100~Mtets. The boundary layers of the flowfield were properly discretized
with elements of considerable stretching. As before, optimal
discretizations have been employed for each field, and the grids do
not match at the interface. Based on previous scoping runs, during each
barely coupled step the flowfield was advanced for 1,500~timesteps.
This number of timesteps was required for the flowfield to reach a quasi 
steady state. The temperature field, on the other hand, was advanced for 
10~timesteps with a timestep of $\Delta t = 10$.
Figures~3.2f-i show the temperature field 
obtained, the velocity in the gap, the velocity of the flowfield and
the particles, as well as the temperature of the flow and the particles.
The run was carried out in mixed OMP/MPI mode on a machine with 8~nodes
of 128~cores each.

\vbox{
\vskip 18pt
\cl{
\includegraphics[width=12.0cm]{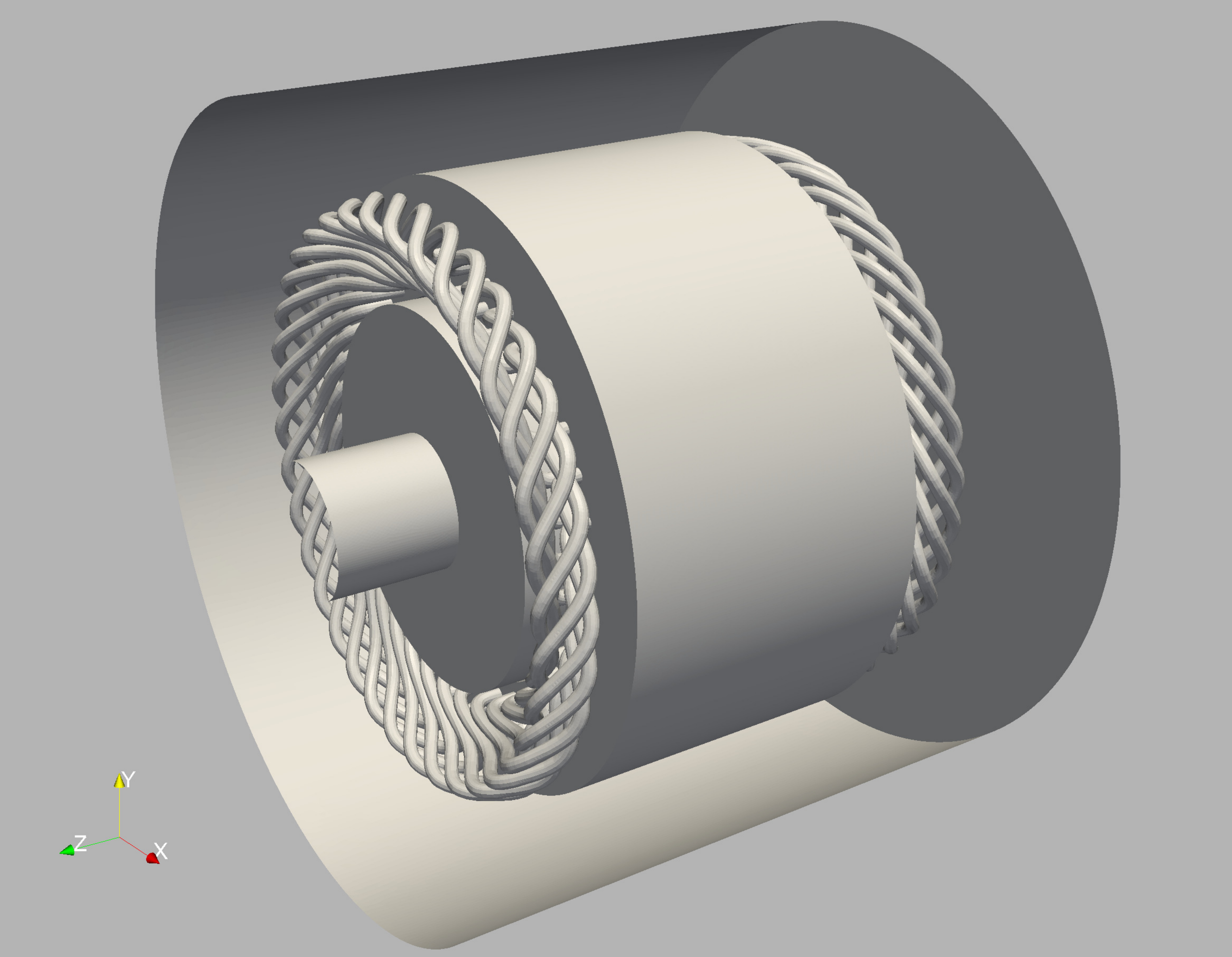}
}
\vskip 5pt
\cl{Figure 3.2a~~E-Motor: Motor in Casing}
}

\vbox{ 
\vskip 18pt
\cl{
\includegraphics[width=12.0cm]{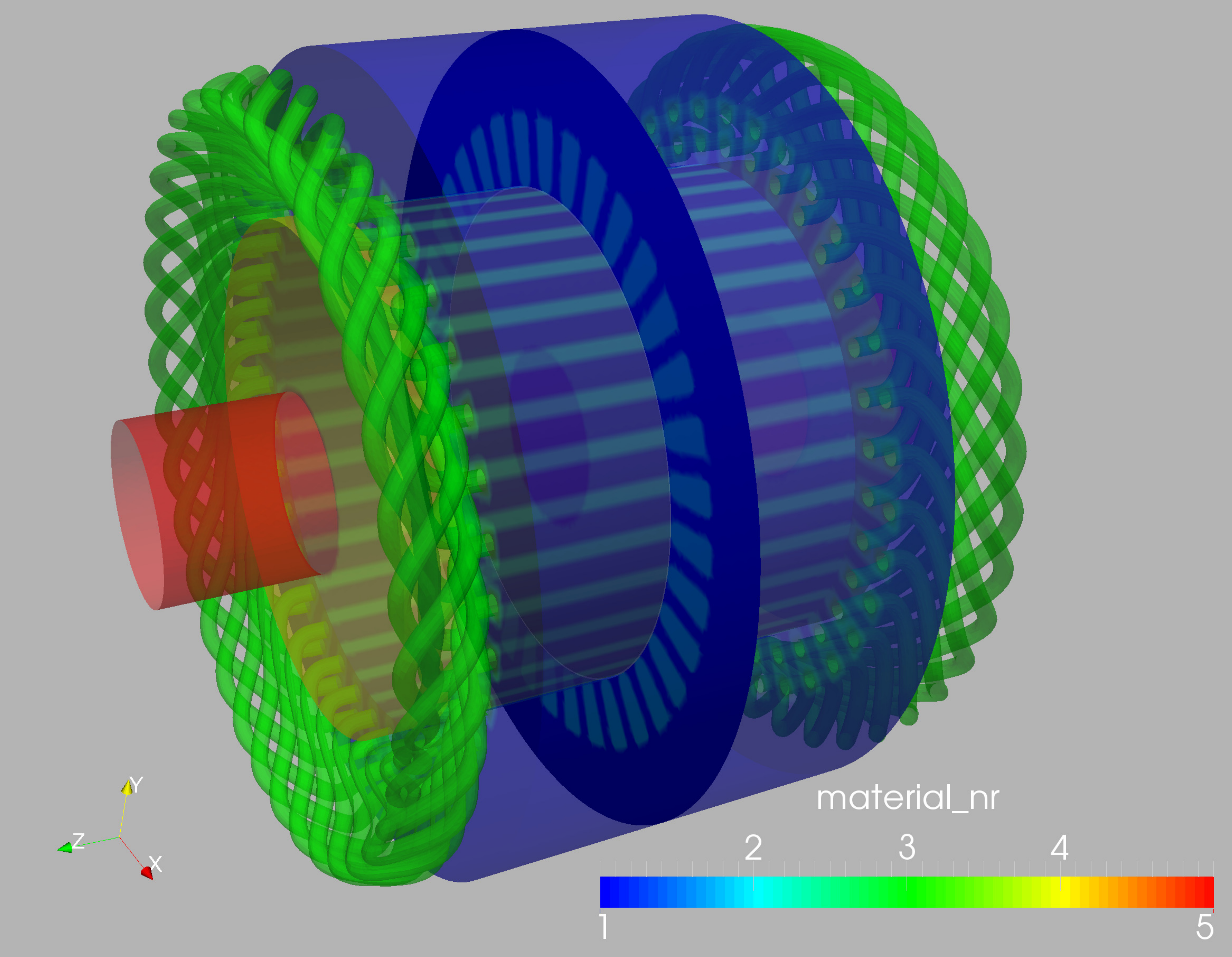}
}
\vskip 5pt
\cl{Figure 3.2b~~E-Motor: Materials Used}
}

\vbox{
\vskip 18pt
\cl{
\includegraphics[width=12.0cm]{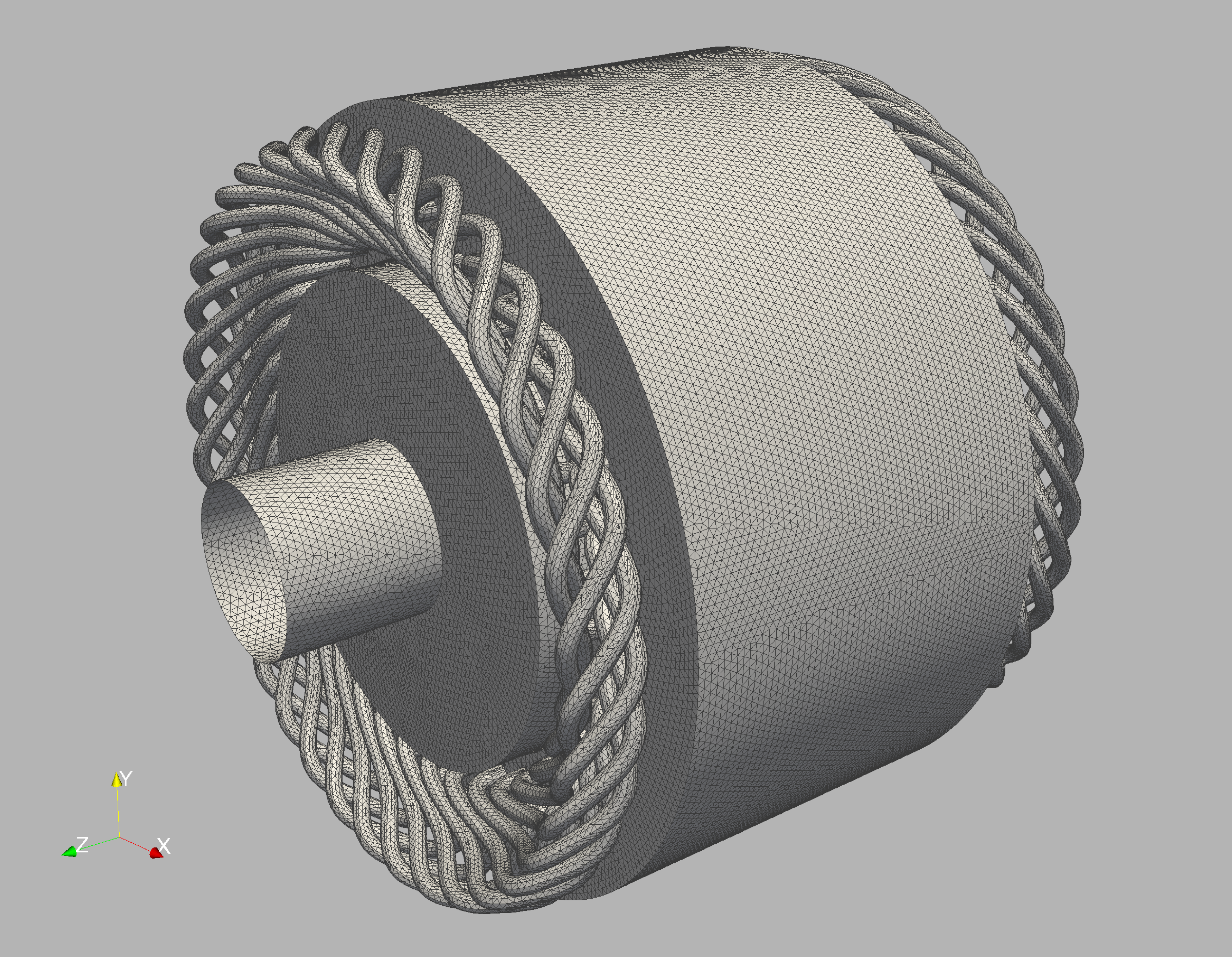}
}
\vskip 5pt
\cl{Figure 3.2c~~E-Motor: Surface Mesh of FEHEAT Region}
}

\vbox{
\vskip 18pt
\cl{
\includegraphics[width=12.0cm]{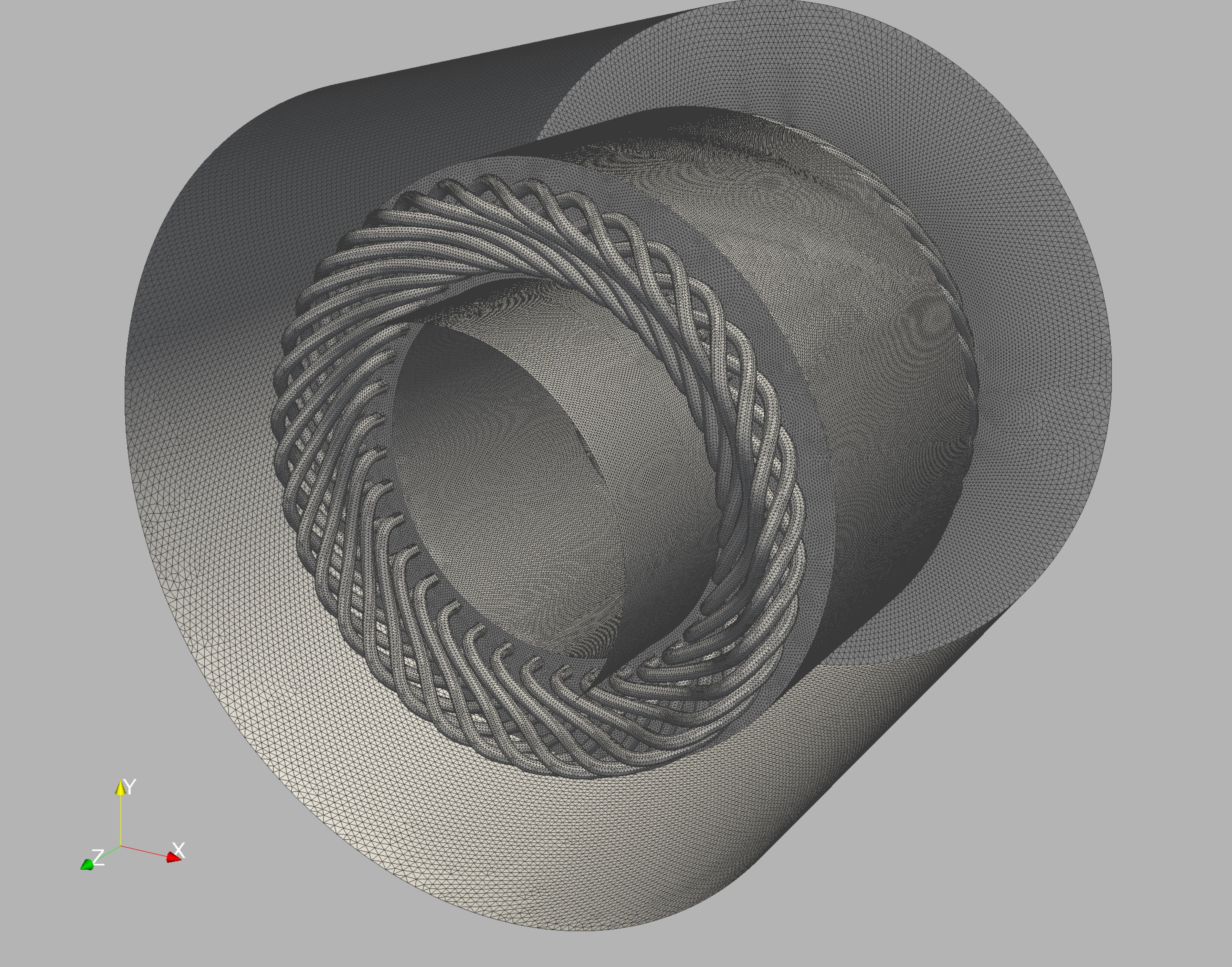}
}
\vskip 5pt
\cl{Figure 3.2d~~E-Motor: Surface Mesh of FEFLO Region}
}

\vbox{
\vskip 18pt
\cl{
\includegraphics[width=12.0cm]{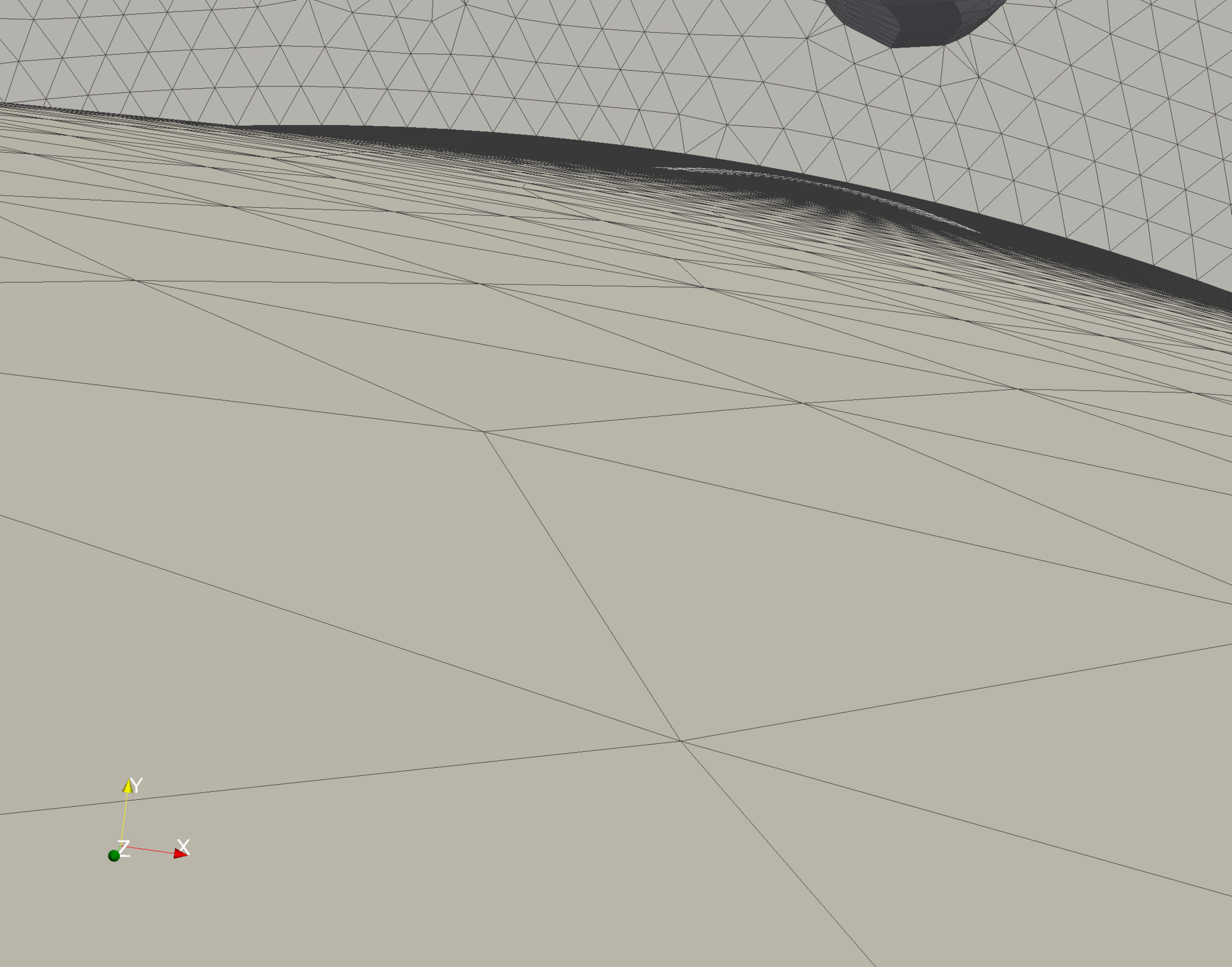}
}
\vskip 5pt
\cl{Figure 3.2e~~E-Motor: Gap Region (Detail)}
}

\vbox{
\vskip 18pt
\cl{
\includegraphics[width=12.0cm]{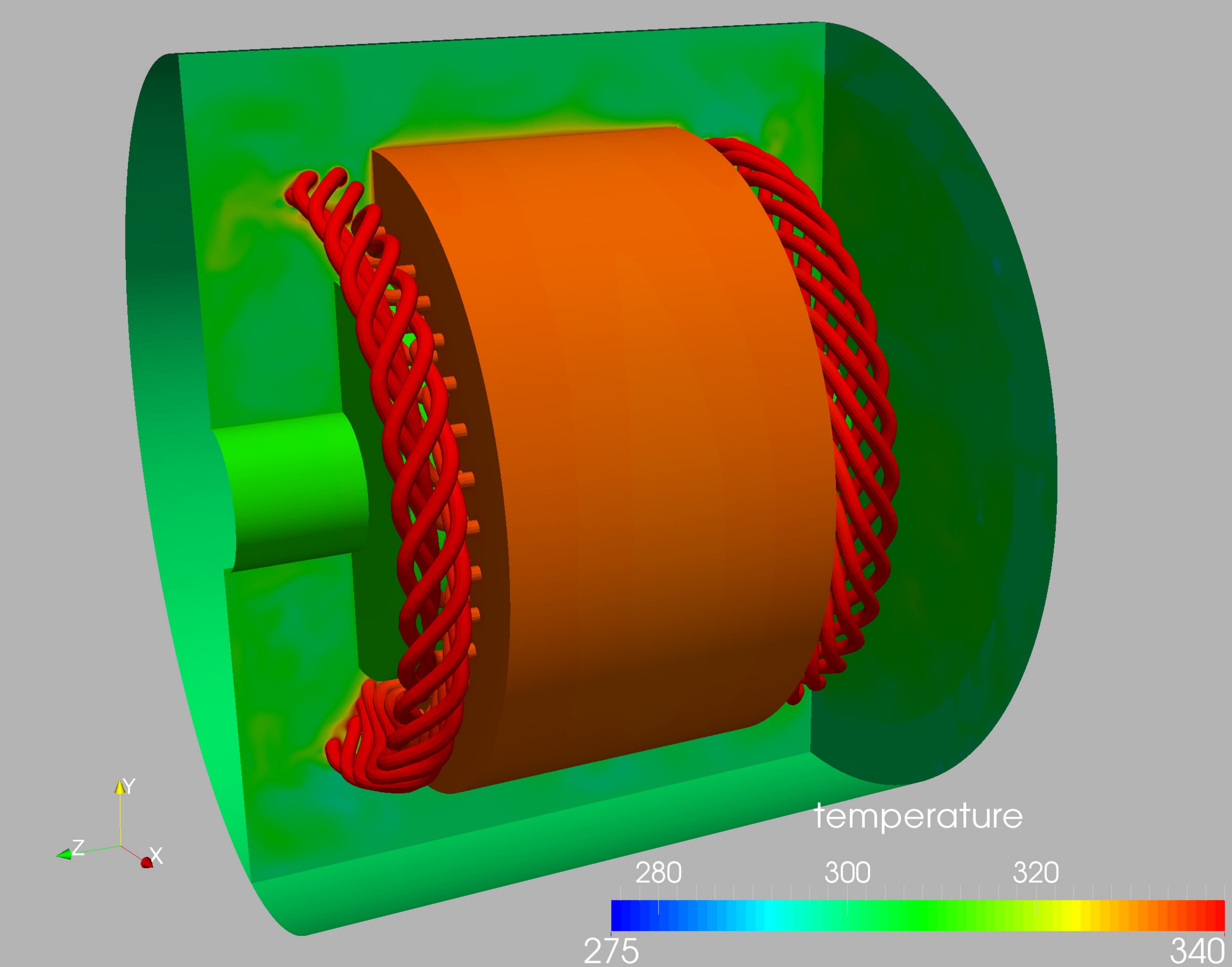}
}
\vskip 5pt
\cl{Figure 3.2f~~E-Motor: Temperature Field}
}

\vbox{     
\vskip 18pt
\cl{
\includegraphics[width=12.0cm]{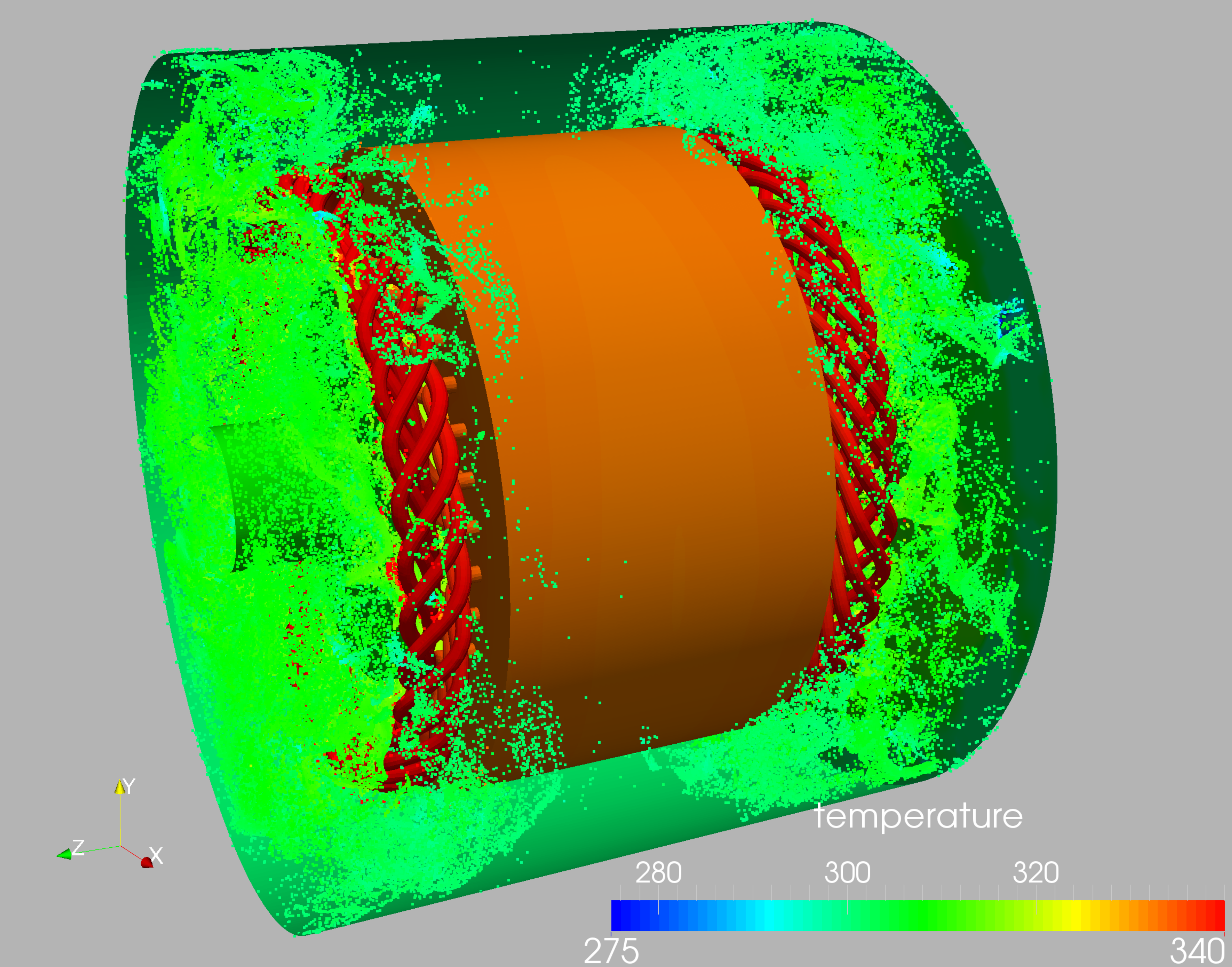}
}
\vskip 5pt
\cl{Figure 3.2g~~E-Motor: Temperature Field for Solid and Particles}
}

\vbox{     
\vskip 18pt
\cl{
\includegraphics[width=12.0cm]{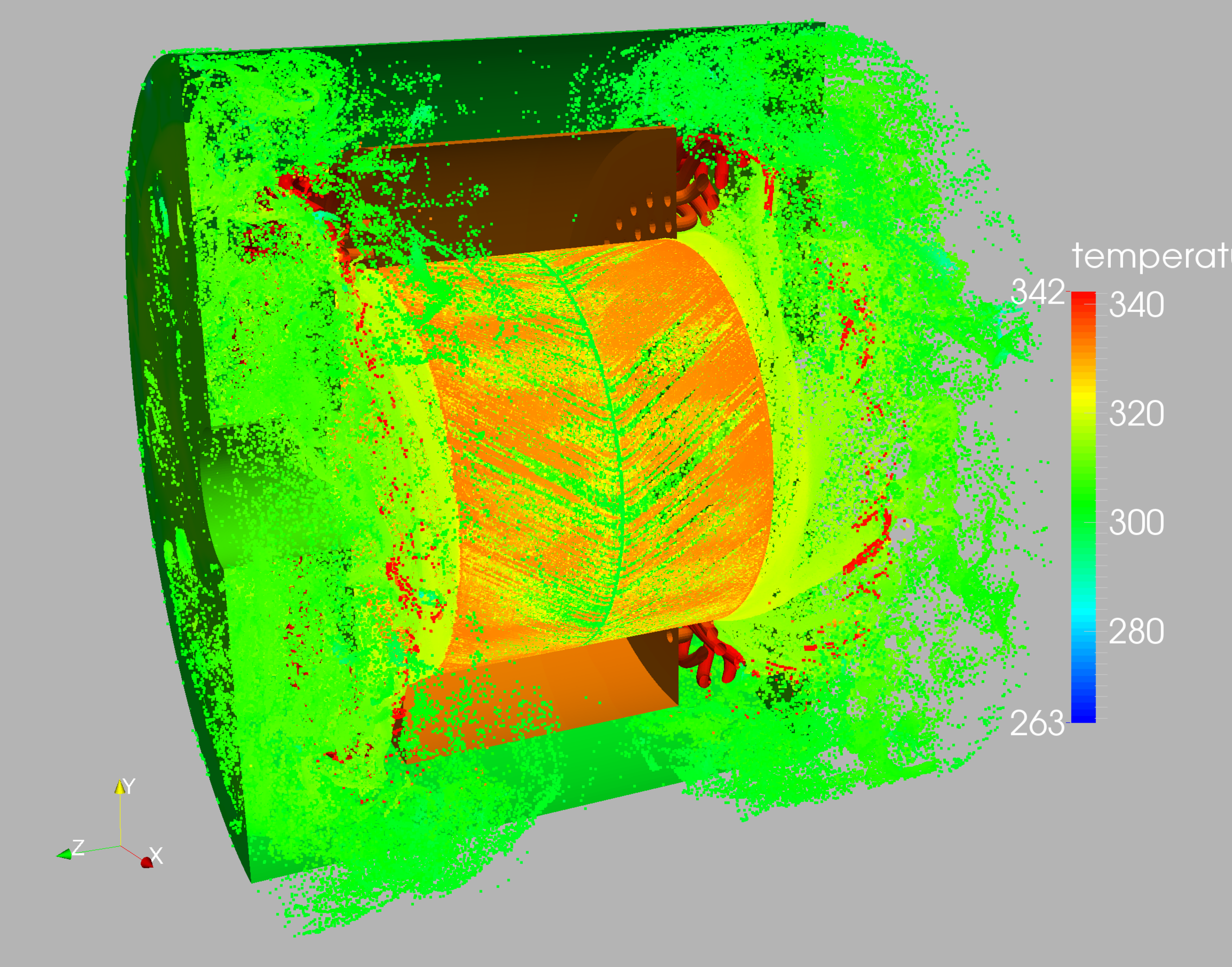}
}
\vskip 5pt 
\cl{Figure 3.2h~~E-Motor: Temperature Field for Solid and Particles}
}

\vbox{
\vskip 18pt
\cl{
\includegraphics[width=8.0cm]{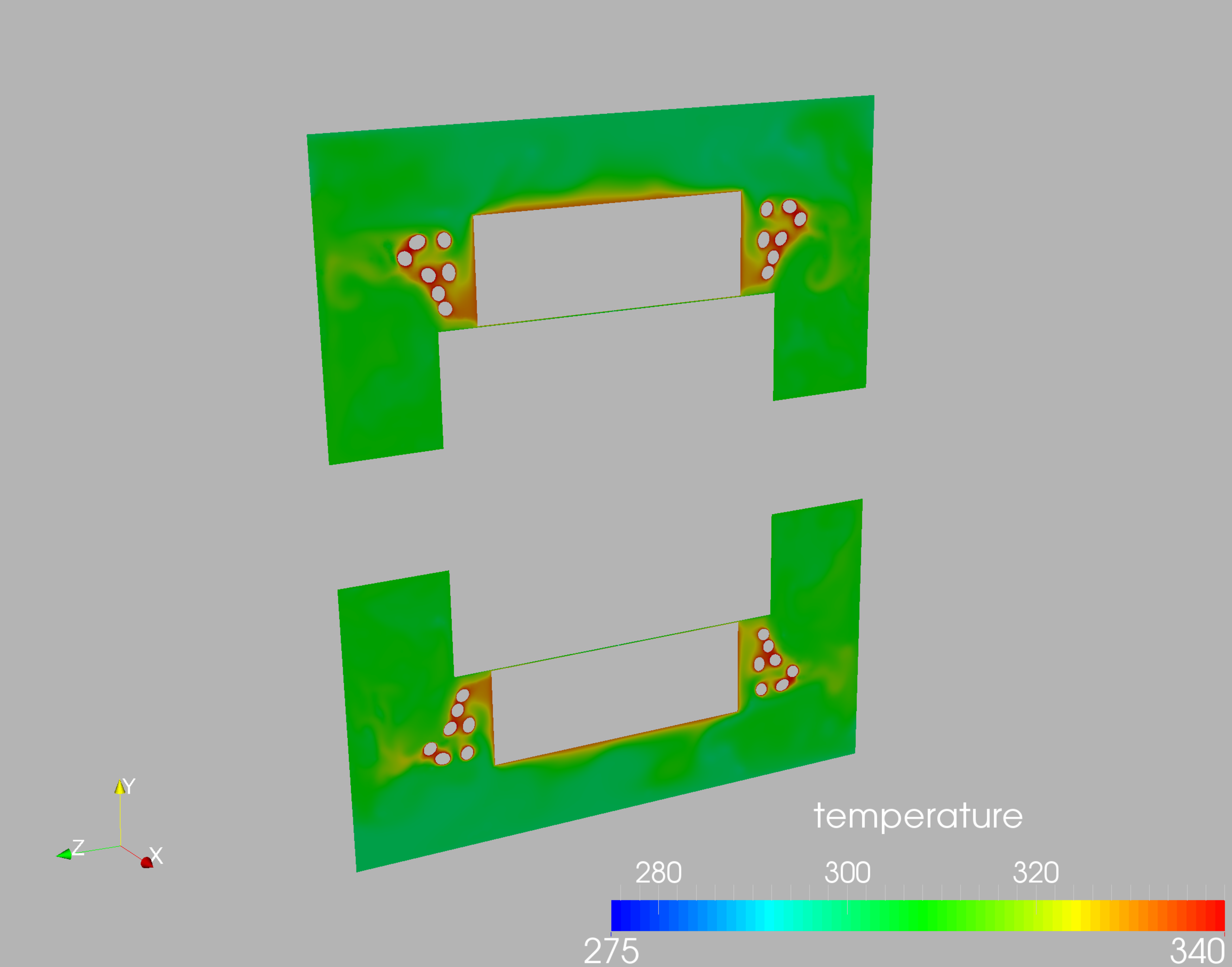}
\hskip 0.2cm
\includegraphics[width=8.0cm]{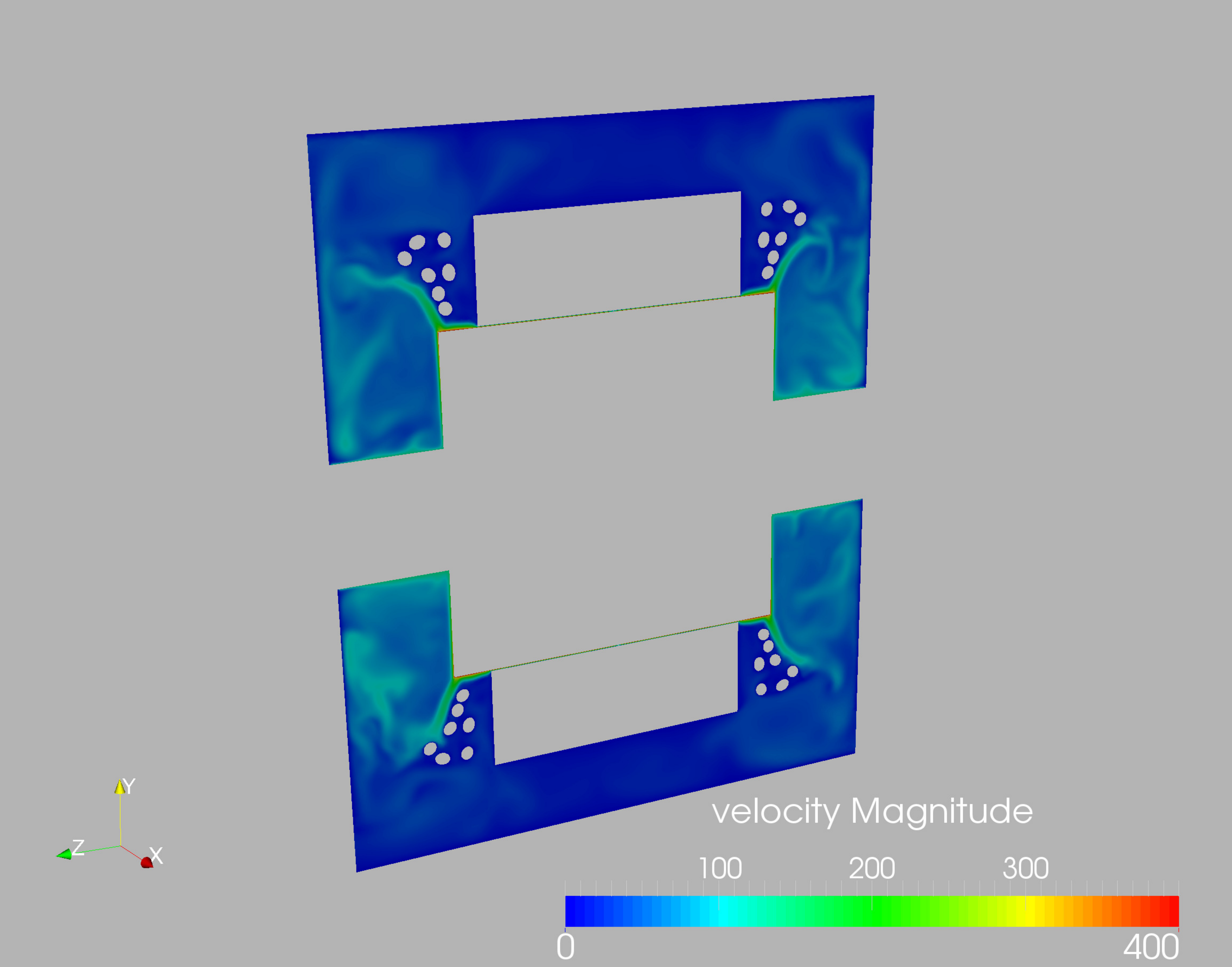}
}
\vskip 5pt
\cl{Figure 3.2i~~E-Motor: Temperature and Velocities in Plane $x=0$}
}


\section{CONCLUSIONS AND OUTLOOK}
A technique to combine codes to solve barely coupled multiphysics problems
has been developed. Each field is advanced separately until a stop is
triggered. This could be due to a preset time increment, a preset number
of timesteps, a preset decrease of residuals, a preset change in unknowns,
a preset change in geometry, or any other physically meaningful quantity.
The technique allows for a simple implementation in coupled codes using
the loose coupling approach. \\
Examples from evaporative cooling of electric motors shows the
viability and accuracy of the proposed procedure. \\
Open questions include:
\begin{itemize}
\item[-] The optimal increase (`ramping up') of time-increments/timesteps
used in the individual discipline/metier codes;
\item[-] The automatic detection of when to stop/exit the individual discipline/metier codes and couple back;
\item[-] The implementation of barely coupled timestepping in optimization 
loops/packages; and
\item[-] The derivation of adjoints for this class of problems (different 
grids, different timescales, etc.).
\end{itemize}


\section{ACKNOWLEDGEMENTS}
This work is partially supported by NSF grant DMS-2408877, the 
Air Force Office of Scientific Research under 
Award NO: FA9550-22-1-0248, and the Office of Naval Research (ONR) under
Award NO: N00014-24-1-2147. \\
This work was also partially supported by the joint DFG/FWF 
Collaborative Research Centre CREATOR (CRC – TRR361 / 10.55776/F90) 
at TU Darmstadt, TU Graz and JKU Linz. We thank JSOL for making 
the JMAG software available to model the geometry.


\bibliographystyle{aiaa}

\newpage

\section{Appendix 1: Physical Modeling of Droplet Propagation and Evaporation}
\label{physmodeldroplets}
When solving the two-phase equations, the air, as a continuum,
is best represented by a set of partial differential equations
(the Navier-Stokes equations) that are numerically solved on a mesh.
Thus, the gas characteristics are calculated
at the mesh points within the flowfield.
The droplets/particles, are modeled using a Lagrangian description, 
where individual particles (or groups of particles) are monitored 
and tracked in the flow, allowing for an exchange of mass, momentum 
and energy between the air and the particles.

\subsection{Equations Describing the Motion of the Air}
As seen from the experimental evidence, the velocities of the air
in the gap region never exceed a Mach-number
of $Ma=0.1$. Therefore, the air may be assumed as a Newtonian,
incompressible liquid. Given the narrow gap and the short timescales,
buoyancy and the effect of gravity may be neglected. The equations 
describing the conservation of momentum, mass and energy for incompressible,
Newtonian flows may be written as

$$ \rho \vvec_{,t} + \rho \vvec \cdot \Grd \vvec + \Grd p = 
 \nabla \cdot \mu \nabla \vvec + \svec_v   ~~, 
                                                         \eqno(A.1.1) $$
$$                             \Div \vvec = 0      ~~,   \eqno(A.1.2) $$
$$ \rho c_p T_{,t} + \rho c_p \vvec \cdot \Grd T = 
       \nabla \cdot k \nabla T + s_e ~~.
                                                         \eqno(A.1.3) $$

\noi
Here $\rho, \vvec, p, \mu, T, c_p, k$ denote
the density, velocity vector, pressure, viscosity, 
temperature, specific heat coefficient and conductivity respectively, and
$\svec_v, s_e$ momentum and energy source terms (e.g. due to particles
or external forces/heat sources). 

\subsection{Equations Describing the Motion of Particles/Droplets}
\label{partmotion}
In order to describe the interaction of particles/droplets with the
flow, the mass, forces and energy/work exchanged between the
flowfield and the particles must be defined.
As before, we denote for {\bf fluid (air)}
by $\rho, p, T, k, v_i, \mu$ and $c_p$ the density, pressure,
temperature, conductivity, velocity in direction $x_i$,
viscosity, and the specific heat at constant pressure.
For the {\bf particles}, we denote by
$\rho_p, T_p, v_{pi}, d, c_{pp}$ and
$Q$ the density, temperature, velocity in direction $x_i$,
equivalent diameter, specific heat coefficient and heat transferred
per unit volume. In what follows we will refer to droplet and
particles collectively as particles. \\
Making the classical assumptions that the particles may be represented
by an equivalent sphere of diameter $d$, the drag forces $\Dvec$
acting on the particles will be due to the difference of fluid and
particle velocity:

$$ \Dvec = {{\pi d^2} \over 4} \cdot c_d \cdot
         { 1 \over 2} \rho | \vvec - \vvec_p | ( \vvec - \vvec_p ) 
                                                  ~~.  \eqno(A.2.1) $$

\noi
The {\bf drag coefficient} $c_d$ is obtained empirically from the
Reynolds-number $Re$:

$$ Re = {{\rho | \vvec - \vvec_p | d } \over { \mu }}  \eqno(A.2.2) $$

\noi
as (see, e.g. \cite{Sch79}):

$$ c_d = max\left(0.1 ,
{24 \over Re} \left( 1 + 0.15 Re^{0.687} \right) \right) \eqno(A.2.3) $$

\noi
The lower bound of $c_d=0.1$ is required to obtain the proper limit for
the Euler equations, when $Re \rightarrow \infty$.
\noi
The heat transferred between the particles and the fluid is given by

$$ Q = {{\pi d^2} \over 4} \cdot 
      \left[ h_f      \cdot ( T   - T_p   )
           + \sigma^* \cdot ( T^4 - T_p^4 ) \right]
                                                 ~~,  \eqno(A.2.4) $$
\noi
where $h_f$ is the film coefficient and $\sigma^*$ the radiation
coefficient. For the class of problems considered here, the particle
temperature and kinetic energy are such that the radiation
coefficient $\sigma^*$ may be ignored. The film coefficient $h_f$ is
obtained from the Nusselt number $Nu$:

$$ Nu = 2 + 0.459 Pr^{0.333} Re^{0.55} ~~,  \eqno(A.2.5)        $$

\noi
where $Pr$ is the Prandtl number of the gas

$$ Pr = {k \over \mu } ~~,  \eqno(A.2.6)         $$

\noi
as

$$ h_f = {{ Nu \cdot k }\over d} ~~.      \eqno(A.2.7)    $$

\par \noi
Having established the forces and heat flux, the particle motion
and temperature are obtained from Newton's law and the first law
of thermodynamics. For the particle velocities, we have:

$$ \rho_p {{\pi d^3} \over 6 } \cdot {{ d\vvec_p} \over {dt}} = \Dvec  ~~.
                                                      \eqno(A.2.8)    $$

\noi
This implies that:

$$ {{ d\vvec_p} \over {dt}} = {{3 \rho} \over {4 \rho_p d}} \cdot c_d
                               | \vvec - \vvec_p | ( \vvec - \vvec_p ) 
                    = \alpha_v | \vvec - \vvec_p | ( \vvec - \vvec_p ) ~~,
                                                      \eqno(A.2.9)   $$

\noi
where $\alpha_v=3\rho c_d / (4 \rho_p d)$.
The particle positions are obtained from:

$$ {{ d\xvec_p} \over {dt}} = \vvec_p ~~.     \eqno(A.2.10)    $$

\noi
The temperature change in a particle is given by:

$$ \rho_p c_{pp} {{\pi d^3} \over 6 } \cdot {{ dT_p} \over {dt}} = Q ~~,
                                                      \eqno(A.2.11) $$

\noi
which may be expressed as:

$$ {{ dT_p} \over {dt}} = {{3 k}\over{2 c_{pp} \rho_p d^2}} \cdot Nu \cdot
                          ( T - T_p )
                        = \alpha_T ( T - T_p ) ~~,    \eqno(A.2.12) $$

\noi
with $\alpha_T=3 k/(2 c_{pp} \rho_p d^2)$.
Equations (A.2.9, A.2.10, A.2.12) may be formulated as a system
of Ordinary Differential Equations (ODEs) of the form:

$$ {{d\uvec_p} \over {dt}} = \rvec(\uvec_p, \xvec, \uvec_f) ~~,
                                                     \eqno(A.2.13)  $$

\noi
where $\uvec_p, \xvec, \uvec_f$ denote the particle unknowns, the
position of the particle and the fluid unknowns at the position of
the particle.

\subsection{Numerical Integration of the Motion of the Air}
\label{numintnavto}
The last six decades have seen a large number of schemes that may be
used to solve numerically the incompressible Navier-Stokes
equations given by Eqns.(A.1.1-A.1.3). In the present case, the
following design criteria were implemented:
\begin{itemize}
\item[-] Spatial discretization using {\bf unstructured grids}
(in order to allow for arbitrary geometries and adaptive refinement);
\item[-] Spatial approximation of unknowns with
{\bf simple linear finite elements} (in order to have a simple
input/output and code structure);
\item[-] Edge-based data structures (for reduced access to memory and
indirect addressing);
\item[-] Temporal approximation using {\bf implicit integration of viscous
terms and pressure} (the interesting scales are the ones associated with
advection);
\item[-] Temporal approximation using {\bf explicit, high-order
integration of advective terms};
\item[-] {\bf Low-storage, iterative solvers} with
{\bf deflation-based preconditioning} \cite{Aub08,Loh11}
for the resulting systems of
equations (in order to solve large 3-D problems); and
\item[-] Steady results that are {\bf independent from the timestep} chosen
(in order to have confidence in convergence studies).
\end{itemize}
\noindent
The resulting discretization in time is given by the following projection
scheme \cite{Loh04,Loh06}:
\begin{itemize}
\item[-] \ub{Advective-Diffusive Prediction}:
$\vvec^n, p^n \rightarrow \vvec^{*}$

$$ \svec' = - \Grd p^n + \rho \gvec 
          + \beta \rho \gvec (T^n - T_0) + \svec_v ~~,
\eqno(A.3.1)
$$

$$
\vvec^i = \vvec^n + \alpha^i \gamma \dt \left(
 - \vvec^{i-1} \cdot \Grd \vvec^{i-1} 
 + \nabla \cdot \mu \nabla \vvec^{i-1} + \svec' \right)  ~~; ~~i=1,k-1~~;
\eqno(A.3.2)
$$

$$
 \left[ { 1 \over \dt} - \theta \nabla \cdot \mu \nabla \right]
   \left( \vvec^{k} - \vvec^n \right)
 + \vvec^{k-1} \cdot \Grd \vvec^{k-1} = 
   \nabla \cdot \mu \nabla \vvec^{k-1} + \svec' ~~.  \eqno(A.3.3)
$$

\ms \noi
\item[-] \ub{Pressure Correction}: $p^n \rightarrow p^{n+1}$

$$
 \Div \vvec^{n+1} = 0                       ~~; \eqno(A.3.4)
$$
$$
 {{ \vvec^{n+1} - \vvec^{*} }\over \dt} + \Grd ( p^{n+1} - p^n )
   = 0                                      ~~; \eqno(A.3.5)
$$

\noi
\item[ ] which results in

$$
 \nabla^2 ( p^{n+1} - p^n ) = {{\Div \vvec^{*} }\over \dt} ~~;
\eqno(A.3.6)
$$

\ms \noi
\item[-] \ub{Velocity Correction}:
$\vvec^{*} \rightarrow \vvec^{n+1}$

$$
 \vvec^{n+1} = \vvec^{*} - \dt \Grd ( p^{n+1} - p^n ) ~~.
\eqno(A.3.7)
$$
\end{itemize}

\noi
$\theta$ denotes the implicitness-factor for the viscous
terms ($\theta=1$: 1st order, fully implicit, $\theta=0.5$: 2nd order,
Crank-Nicholson).
$\alpha^i$ are the standard low-storage Runge-Kutta coefficients
$\alpha^i=1/(k+1-i)$. The $k-1$ stages of Eqn.(A.3.2) may be seen as a
predictor (or replacement)
of $\vvec^n$ by $\vvec^{k-1}$. The original right-hand side has not been
modified, so that at steady-state $\vvec^n=\vvec^{k-1}$, preserving the
requirement that the steady-state be independent of the timestep $\dt$.
The factor $\gamma$ denotes the local ratio of the stability limit for
explicit timestepping for the viscous terms versus the timestep chosen.
Given that the advective and viscous timestep limits are proportional to:

$$ \dt_a \approx {h \over {|\vvec|}} ~~;~~
   \dt_v \approx {{\rho h^2} \over \mu} ~~, \eqno(A.3.8)
$$

\noi
we immediately obtain

$$ \gamma = {{\dt_v} \over {\dt_a}}
    \approx {{\rho |\vvec| h }\over{\mu}} \approx Re_h  ~~,
\eqno(A.3.9)
$$

\noi
or, in its final form:

$$ \gamma = min(1,Re_h) ~~. \eqno(A.3.10) $$

\noi
In regions away from boundary layers, this factor is $O(1)$, implying
that a high-order Runge-Kutta scheme is recovered. Conversely, for
regions where $Re_h=O(0)$, the scheme reverts back to the usual
1-stage Crank-Nicholson scheme.
Besides higher accuracy, an important benefit of explicit multistage
advection schemes is the larger timestep one can employ. The increase in
allowable timestep is roughly proportional to the number of stages used
(and has been exploited extensively for compressible flow simulations
\cite{Jam81}).
Given that for an incompressible solver of the projection type
given by Eqns.(A.3.1-A.3.7) most of the CPU time is spent solving the
pressure-Poisson system Eqn.(A.3.6), the speedup
achieved is also roughly proportional to the number of stages used. \\
At steady state, $\vvec^{*}=\vvec^n=\vvec^{n+1}$ and the residuals of
the pressure correction vanish,
implying that the result does not depend on the timestep $\dt$. \\
The spatial discretization of these equations is carried out via
linear finite elements. The
resulting matrix system is re-written as an edge-based solver, allowing
the use of consistent numerical fluxes to stabilize the advection and
divergence operators \cite{Loh08}. \\
The energy (temperature) equation (Eqn.(A.3.3)) is integrated in a
manner similar to the advective-diffusive prediction (Eqn.(A.3.2)),
i.e. with an explicit, high order Runge-Kutta scheme for the advective
parts and an implicit, 2nd order Crank-Nicholson scheme for the
conductivity.

\subsection{Numerical Integration of the Motion of Particles/Droplets}
\label{numintpast}
The equations describing the position, velocity and temperature of a
particle (Eqns.\ A.2.9, A.2.10, A.2.12) may be formulated as a
system of nonlinear Ordinary Differential Equations of the form:

$$ {{d\uvec_p} \over {dt}} = \rvec(\uvec_p, \xvec, \uvec_f) ~~.
                                                       \eqno(A.4.1) $$

\par \noi
They can be integrated numerically in a variety of ways. Due to its
speed, low memory requirements and simplicity, we have chosen
the following k-step low-storage Runge-Kutta procedure to integrate them:

$$ \uvec^{n+i}_p = \uvec^n_p + \alpha^i \Delta t \cdot
   \rvec(\uvec^{n+i-1}_p, \xvec^{n+i-1}, \uvec^{n+i-1}_f) ~~,
~~ i=1,k  ~~. \eqno(A.4.2) $$

\noi
For linear ODEs the choice

$$ \alpha^i= {1 \over {k+1-i}} ~~,~~ i=1,k  \eqno(A.4.3) $$

\noi
leads to a scheme that is $k$-th order accurate in time.
Note that in each step the location of the particle with respect to the
fluid mesh needs to be updated in order to obtain the proper values for
the fluid unknowns. The default number of stages used is $k=4$. This
would seem unnecessarily high, given that the flow solver is of
second-order accuracy, and that the particles are integrated separately
from the flow solver before the next (flow) timestep, i.e. in a staggered
manner. However, it was found that the 4-stage particle integration
preserves very well the motion in vortical structures and leads to less
`wall sliding' close to the boundaries of the domain \cite{Loh14}.
The stability/ accuracy of the particle integrator should not be a problem
as the particle motion will always be slower than the maximum wave speed
of the fluid (fluid velocity). \\
The transfer of forces and heat flux between the fluid and the particles
must be accomplished in a conservative way, i.e. whatever is added to the
fluid must be subtracted from the particles and vice-versa. The finite
element discretization of the fluid equations will lead to
a system of ODE's of the form:

$$ \Mmat \duvec = \rvec ~~,    \eqno(A.4.4)       $$

\noi
where $\Mmat, \duvec$ and $\rvec$ denote, respectively, the consistent
mass matrix, increment of the unknowns vector and right-hand side vector. 
Given the `host element' of each particle, i.e. the fluid mesh element
that contains the particle, the forces and heat transferred
to $\rvec$ are added as follows:

$$ \rvec^i_D = \sum_{el~surr~i} N^i(\xvec_p) \Dvec_p ~~.  \eqno(A.4.5) $$

\noi
Here $N^i(\xvec_p)$ denotes the shape-function values of the host
element for the point coordinates $\xvec_p$, and the sum extends
over all elements that surround node $i$. As the sum of all
shape-function values is unity at every point:

$$ \sum N^i(\xvec) = 1 ~~\forall \xvec ~~,    \eqno(A.4.6)       $$

\noi
this procedure is strictly conservative. \\
From Eqns.(A.2.9, A.2.10, A.2.12) and their equivalent numerical
integration via Eqn.(A.4.2),
the change in momentum and energy for one particle is given by:

$$ \fvec_p =  \rho_p {{\pi d^3}\over 6} 
             {{\left( \vvec^{n+1}_p - \vvec^n_p \right)} \over {\Delta t}}
                                           ~~,    \eqno(A.4.7)       $$

$$ q_p =  \rho_p c_{pp} {{\pi d^3}\over 6} 
             {{\left( T^{n+1}_p - T^n_p \right)} \over {\Delta t}}
                                           ~~.    \eqno(A.4.8)       $$

\noi
These quantities are multiplied by the number of particles in
a packet in order to obtain the final values transmitted to the fluid.
Before going on, we summarize the basic steps required in order to update
the particles one timestep:
\begin{itemize}
\item[-] Initialize Fluid Source-Terms: $\rvec=0$
\item[-] {\code DO}: For Each Particle:
\item[ ] - {\code DO}: For Each Runge-Kutta Stage:
\item[ ] ~~~- Find Host Element of Particle: {\code IELEM}, $N^i(\xvec)$
\item[ ] ~~~- Obtain Fluid Variables Required
\item[ ] ~~~- Update Particle: Velocities, Position, Temperature, ...
\item[-] - {\code ENDDO}
\item[ ] - Transfer Loads to Element Nodes
\item[-] {\code ENDDO}
\end{itemize}

\subsubsection{Particle Parcels}
\label{partparc}
For a large number of very small particles, it becomes impossible to
carry every individual particle in a simulation. The solution is to:
\begin{itemize}
\item[a)] Agglomerate the particles into so-called packets of $N_p$
particles;
\item[b)] Integrate the governing equations for one individual particle;
and
\item[c)] Transfer back to the fluid $N_p$ times the effect of one
particle.
\end{itemize}
Beyond a reasonable number of particles per element (typically $> 8$),
this procedure produces accurate results without any deterioration in
physical fidelity \cite{Loh14}.

\subsubsection{Other Particle Numerics}
In order to achieve a robust particle integrator, a number of additional
precautions and algorithms need to be implemented. The most
important of these are:
\begin{itemize}
\item[-] Agglomeration/Subdivision of Particle Parcels:
As the fluid mesh may be adaptively refined and coarsened in time,
or the particle traverses elements of different sizes,
it may be important to adapt the parcel concentrations as well.
This is necessary to ensure that there is sufficient parcel
representation in each element and yet, that there are not too many
parcels as to constitute an inefficient use of CPU and memory.
\item[-] Limiting During Particle Updates:
As the particles are integrated independently from the flow solver, it is
not difficult to envision situations where for the extreme cases of
very light or very heavy particles physically meaningless or unstable
results may be obtained.
In order to prevent this, the changes in
particle velocities and temperatures are limited in order not to exceed
the differences in velocities and temperature between the particles and
the fluid \cite{Loh14}.
\item[-] Particle Contact/Merging:
In some situations, particles may collide or merge in a certain region
of space.
\item[-] Particle Tracking:
A common feature of all particle-grid applications is that the particles
do not move far between timesteps. This makes physical sense:
if a particle jumped ten gridpoints during one timestep, it would have
no chance to exchange information with the points along the way, leading
to serious errors. Therefore, the assumption that the new host elements
of the particles are in the vicinity of the current ones is a valid one.
For this reason, the most efficient way to search for the new host
elements is via the vectorized neighbour-to-neighbour algorithm
described in \cite{Loh08}.
\end{itemize}

\subsection{FEFLO}
\label{FEFLO}
The design criteria and numerical techniques described above were
implemented in FEFLO~\cite{Loh24a}, a general-purpose computational 
fluid dynamics
(CFD) code. The code has had a long history of relevant applications
in incompressible flows {\cite{Ram93,Ram99,Loh04,Loh06,Til08}},
free-surface hydrodynamics {\cite{Yan06}} and
dispersion {\cite{Cam04}}, and
has been ported to vector, shared memory {\cite{Sha00}},
distributed memory {\cite{Ram93,Ram96}} and GPU-based
\cite{Cor12} machines.

\newpage

\section{Appendix 2: Modeling of Heat Propagation in the Motor}
\label{physmodelheat}

\subsection{Equation Describing the Temperature in the Solid}
The equation describing the conservation energy for solids
may be written as

$$ \rho c_p T_{,t} = \nabla \cdot k \nabla T + s_m ~~.  \eqno(B.1.3) $$

\noi
Here $\rho, T, c_p, k$ denote the density, temperature, specific heat 
coefficient and conductivity respectively, and
$\svec_m$ the source terms (e.g. due to magnetic hysteresis).

\subsection{Numerical Integration of the Heat Equation}
\label{numintheat}
For the numerical integration of the heat equation the 
following design criteria were implemented:
\begin{itemize}
\item[-] Spatial discretization using {\bf unstructured grids}
(in order to allow for arbitrary geometries and adaptive refinement);
\item[-] Spatial approximation of unknowns with
{\bf simple linear finite elements} (in order to have a simple
input/output and code structure);
\item[-] Edge-based data structures (for reduced access to memory and
indirect addressing);
\item[-] Temporal approximation using {\bf implicit integration};
\item[-] {\bf Low-storage, iterative solvers} with
{\bf deflation-based preconditioning} \cite{Aub08,Loh11} for the 
resulting systems of
equations (in order to solve large 3-D problems).
\end{itemize}
\noindent
The resulting discretization in time is given by the usual $\Theta$ scheme:

$$
 \left[ { {\rho c_p} \over \dt} - \Theta \nabla \cdot k \nabla \right]
   \Delta T = 
   \nabla \cdot k \nabla T + \svec ~~.  \eqno(B.2.1)
$$

\subsection{FEHEAT}
\label{FEHEAT}
The design criteria and numerical techniques described above were
implemented in FEHEAT \cite{Loh24b}, a general-purpose computational 
thermo dynamics
(CTD) code. The code has had a long history of relevant applications
in heat conduction and aerothermodynamics {\cite{Loh94,Loh98}},
and has been ported to vector, shared memory, distributed memory
and GPU-based machines.

\end{document}

%% file: abbrevs.tex
\font\code=cmtt10

\def\pmb#1{\setbox0=\hbox{$#1$}%
             \kern-.027em\copy0\kern-\wd0
             \kern+.009em\copy0\kern-\wd0
             \kern+.009em\copy0\kern-\wd0
             \kern+.009em\copy0\kern-\wd0
             \kern+.009em\copy0\kern-\wd0
             \kern+.009em\copy0\kern-\wd0
             \kern+.009em\copy0\kern-\wd0
             \kern-.045em\raise+.012em\copy0\kern-\wd0
             \kern+.009em\raise+.012em\copy0\kern-\wd0
             \kern+.009em\raise+.012em\copy0\kern-\wd0
             \kern+.009em\raise-.012em\copy0\kern-\wd0
             \kern+.009em\raise-.012em\copy0\kern-\wd0
             \kern-.018em\copy0\kern-\wd0\raise-.012em\box0}
\def\Pmb#1{\setbox0=\hbox{$#1$}%
             \kern-.033em\copy0\kern-\wd0
             \kern+.011em\copy0\kern-\wd0
             \kern+.011em\copy0\kern-\wd0
             \kern+.011em\copy0\kern-\wd0
             \kern+.011em\copy0\kern-\wd0
             \kern+.011em\copy0\kern-\wd0
             \kern+.011em\copy0\kern-\wd0
             \kern-.055em\raise+.015em\copy0\kern-\wd0
             \kern+.011em\raise+.015em\copy0\kern-\wd0
             \kern+.011em\raise+.015em\copy0\kern-\wd0
             \kern+.011em\raise-.015em\copy0\kern-\wd0
             \kern+.011em\raise-.015em\copy0\kern-\wd0
             \kern-.022em\copy0\kern-\wd0\raise-.015em\box0}
\def\cl{\centerline}

\def\ms{\medskip}
\def\bs{\bigskip}
\def\ub{\underbar}
\def\noi{\noindent}

\def\fvec{{\bf f}}
\def\gvec{{\bf g}}

\def\rvec{{\bf r}}
\def\svec{{\bf s}}

\def\uvec{{\bf u}}
\def\vvec{{\bf v}}

\def\xvec{{\bf x}}

\def\Dvec{{\bf D}}

\def\Mmat{{\bf M}}

\def\dt{{\Delta t}}

\def\Grd{\nabla}
\def\Div{\nabla \cdot}

\def\duvec{\Delta {\bf u}}

\vfill